\documentclass[aps,twocolumn,superscriptaddress,nofootinbib]{revtex4}

\usepackage{amsmath}
\usepackage{amsfonts}
\usepackage{amssymb}
\usepackage{mathdots}

\usepackage{graphicx}
\usepackage{bbm}
\usepackage{color}
\usepackage{pifont}
\usepackage{enumerate}
\usepackage{subfigure}

\DeclareMathOperator{\tr}{tr}

\newcommand{\ket}[1]{|#1\rangle}
\newcommand{\bra}[1]{\langle#1|}

\newcommand{\ketbra}[2]{\ket{#1}\bra{#2}}
\newcommand{\braket}[2]{\langle #1 | #2 \rangle}

\newcommand{\e}{\mathrm{e}}
\newcommand{\startproof}{\textit{Proof. }}
\newcommand{\qedendproof}{\hfill \ensuremath{\blacksquare}}

\newcommand{\rh}[1]{\rho_{#1}}
\newcommand{\Gm}{\mathbf \Gamma}
\newcommand{\rhoAB}{\rho_{AB}}
\newcommand{\sigABC}{\sigma_{ABC}}

\newtheorem{theorem}{Theorem}
\newtheorem{lemma}{Lemma}

\newtheorem{definition}{Definition}
\newtheorem{observation}{Observation}

\begin{document}


\title{Symmetries in Quantum Key Distribution and the Connection between Optimal Attacks and Optimal Cloning}


\author{Agnes Ferenczi}
\email[]{aferenczi@iqc.ca}
\affiliation{Institute for Quantum Computing \& Department for Physics and Astronomy, University of Waterloo, 200 University Avenue West, N2L 3G1, Waterloo, Ontario, Canada }
\author{ Norbert L\" utkenhaus}
\affiliation{Institute for Quantum Computing \& Department for Physics and Astronomy, University of Waterloo, 200 University Avenue West, N2L 3G1, Waterloo, Ontario, Canada }



\date{\today}

\begin{abstract}

We investigate the connection between the optimal collective eavesdropping attack and the optimal cloning attack where the eavesdropper employs an optimal cloner to attack the quantum key distribution (QKD) protocol. The analysis is done in the context of the security proof in Refs. \cite{devetak05a, kraus05a} for discrete variable protocols in $d$-dimensional Hilbert spaces. 
We consider a scenario in which the protocols and cloners are equipped with symmetries. These symmetries are used to define a quantum cloning scenario. 
We find that, in general, it does not hold that the optimal attack is an optimal cloner. However, there are classes of protocols, where we can identify an optimal attack by an optimal cloner. 
We analyze protocols with $2$, $d$ and $d+1$ mutually unbiased bases where $d$ is a prime, and show that for the protocols with $2$ and $d+1$ MUBs the optimal attack is an optimal cloner, but for the protocols with $d$ MUBs, it is not \footnote{These results were already presented in the Poster session at QIP 2010, Zurich, Switzerland}. 
Finally, we give criteria to identify protocols which have different signal states, but the same optimal attack. Using these criteria, we present qubit protocols which have the same optimal attack as the BB84 protocol or the 6-state protocol.

\end{abstract}


\maketitle

\section{Introduction}  \label{introduction_sec}

The objective of quantum key distribution (QKD) is to establish a secret key between two legitimate parties (Alice and Bob), that is unknown to an eavesdropper (Eve). The secret key can be used later in cryptographic applications, for example to facilitate secure communication.

To start a QKD protocol  Alice prepares quantum states (signal states) and sends them through a quantum channel to Bob, who performs measurements on them. These type of protocols are referred to as prepare-and-measure protocols and result in quantum mechanically correlated classical data being shared between Alice and Bob. They can extract a secret key from this data using classical communication protocols. For this to succeed, Alice and Bob need to be able to upper bound the amount of information Eve can gain on the correlated data. This information comes from an interaction of Eve with the signals. For any such attack, there is a trade-off between the amount of information that leaks to Eve and the amount of disturbance that she causes to the signal states. From the observation of this disturbance, Alice and Bob can estimate Eve's information on the data and perform suitable communication protocols to distill secret keys form their data on which Eve has no information.

In this paper, we deal with the problem of finding the optimal interaction between Eve and the signals. Our approach to the security analysis is to restrict Eve to collective attacks \cite{devetak05a, kraus05a} in which she interacts with each signal separately as the range of the validity of this attack can be shown to extend to the most general attack (see section \ref{key_sec}). A collective attack is completely determined by a unitary interaction $U_E$ between the signal states and some additional ancilla states held by Eve. The optimal attack which gives the highest amount of information to Eve (by some suitable measure) for a given amount of disturbance is denoted by $U_E^{\mathrm{opt}}$.

One specific type of interaction is an optimal quantum cloner \cite{buzek96a}. An optimal cloner is a unitary transformation $U_C^{\mathrm{opt}}$ that acts on the signal states and some ancilla states, with the objective of producing two copies of the signal states. The optimal cloner has the property that the copies emerge with the highest fidelity (with respect to the original signal states) allowed by quantum mechanics. An optimal cloner is called symmetric if the fidelities of the two copies are the same, and asymmetric if the fidelities are different. 
Consider now the following eavesdropping attack:
Eve uses an optimal asymmetric cloner to copy the signal states sent by Alice, forwards one copy to Bob, and keeps the other copy for herself. 
She choses the optimal cloner in such a way that the fidelity of Bob's copy is in agreement with Bob's measurement outcomes.
In \cite{cerf02a, durt03a, durt04a}, such cloning attacks were used to model Eve's attack, but optimality was only conjectured. Indeed, for some protocols (e.g. the BB84 \cite{bennett84a} or the 6-state protocol \cite{bruss98a}), the optimal attack is known to be an optimal cloner, but in general the relationship between optimal cloning and optimal eavesdropping is unknown. 

The goal of the present work is to establish the connection between optimal eavesdropping on QKD protocols and optimal cloning in the context of the security definition in Refs. \cite{devetak05a, kraus05a} for protocols with direct, one-way reconciliation. We consider protocols with symmetries so that, without loss of generality, the optimal attack is found in a set with a corresponding symmetry. In this scenario, it turns out that a necessary condition for an optimal cloner to be a candidate for an optimal attack is the strong covariance condition \cite{chiribella05a}. This condition ensures that the optimal cloner and the optimal attack are drawn from the same set, and that the optimal cloner uses the same number of ancilla states as the optimal eavesdropping attack. If strong covariance does not hold for the optimal cloner defined by the signal states of the QKD protocol, we can already conclude that the optimal attack on the QKD protocol is not an optimal cloner. 

In this paper, we calculate the optimal attack for qubit-based protocols (e.g. the BB84  and the 6-state protocol) and for protocols in d-dimensional Hilbert spaces using mutually unbiased bases (MUBs) (e.g. protocols with 2 MUBs,  d+1 MUBs \cite{cerf02a} or d MUBs) and compare the results to the optimal cloner. The security of these protocols has been studied previously in Refs. \cite{bruss98a, bechmann99a, fuchs97a,  bruss02a, cerf02a, durt04a}. In Ref. \cite{sheridan10a}, in particular, the security was proven for protocols with 2 and $d+1$ MUBs using the security proof methods of Refs. \cite{devetak05a, kraus05a}. 

Additionally, we observe that some groups of QKD protocols that share some common symmetry features can be proven to have the same optimal attack, despite having different sets of signal states. As an example, we present qubit protocols which have the same optimal attack as the BB84 protocol or the 6-state protocol, and we give criteria to identify protocols with the same optimal attack. 

This paper is organized as follows.
In section \ref{source_sec} we describe prepare-and-measure protocols using a thought set-up that includes entangled states and is referred to as the source-replacement scheme. In section \ref{key_sec} we summarize the security proofs given in Refs.  \cite{devetak05a, kraus05a}. In section \ref{theorems_sec}, we provide two theorems about the convexity and concavity properties, as well as a lemma about the invariance property of the classical mutual information and the Holevo quantity. In section \ref{symmetries_sec} we describe protocols with a symmetry in the signal states.
Under the assumption that Alice and Bob use only averaged measurement quantities, and that the classical post processing respects certain symmetry properties, we show that the symmetries of the signal states can be transfered to Eve's interaction. This holds true in particular for the class of protocols where the signal states are composed of complete sets of basis states, and where Alice and Bob discard data where they disagree on the basis after the measurement.
The formalism of quantum cloners is summarized in section \ref{cloners_sec} along the lines of Ref. \cite{chiribella05a}. 
The main statement of section \ref{connection_sec} is that the optimal cloner must be strong covariant in order to be an optimal attack. However, strong covariance alone does not yet uniquely determine if the optimal attack is an optimal cloner. We summarize the features of the class of protocol where the signal states are invariant under the generalized Pauli group, for which the corresponding cloning attack is known to be strong covariant.
In section \ref{examples_sec} we give examples of protocols which use the mutually unbiased eigenbases of the generalized Pauli operators and analyze the relation between the optimal attack and the optimal cloner.
In section \ref{classes_sec} we provide a theorem for the class of protocols with complete sets of basis states and basis sifting, which allows us to determine when the optimal attacks of different protocols in this class are the same. Finally, we draw conclusions in section \ref{conclusion_sec}.


\section{Source-replacement scheme} \label{source_sec}

A typical QKD protocol consists of a quantum and a classical phase. In the quantum phase, Alice chooses signal states $\ket{\varphi_x}$ from a set $\mathbf S$ with probability $p(x)$ defined on a $d$-dimensional Hilbert space. She sends the signal states through a quantum channel to Bob, who performs measurements on them by means of a positive operator valued measures (POVM) $\mathbf M_B = \{ B_y \} $, resulting in a joint probability distribution $p(x,y)$. This signal preparation scheme is typically called prepare-and-measure scheme. 
In the classical phase, Alice and Bob perform error correction and privacy amplification, in order extract a secret key from their measurement data. 

The security proof of a protocol is more conveniently described in the source-replacement scheme, which is equivalent to the prepare-and-measure scheme. The source-replacement scheme is a thought set-up, in which Alice creates the bipartite entangled state (source state) 
\begin{equation}
\ket{\Phi}=\sum_x \sqrt{p(x)} \ket{x}_X \ket{\varphi_x}_S \;  \label{source}
\end{equation}
in her lab, keeps the first half for herself and sends the other half to Bob. The states $\ket x$ form an orthonormal basis $\mathcal X=\left \{ \ket{x},x=0,...,|\mathbf S|-1 \right \}$ of an $|\mathbf S|$-dimensional Hilbert space $\mathcal H_X$. In order to prepare the state $\ket{\varphi_x}$ at Bob's side, Alice performs a projective measurement in the basis $\mathcal X$, which triggers the source state to collapse onto the conditional state $\ket{\varphi_x}$ with probability $p(x)$. 

If the signal states are linearly dependent, we can rewrite the source state in a more compact form using the Schmidt decomposition of pure states. For this purpose, we define a $d$-dimensional subsystem $\mathcal H_A$ of $\mathcal H_X$, and express the source state on the ``compressed" space $\mathcal H_A \otimes \mathcal H_S$ 
\begin{equation}
\ket \Phi_{AS} = \sum_{i=0}^{d-1}\sqrt{\kappa_i} \ket{\bar i}_A \ket{i}_S \label{compressed_source}.
\end{equation}
In this expression the Schmidt basis $\mathcal B = \{\ket{i}_S ; i=0,...,d-1 \}$ of system $S$ and the Schmidt coefficients $\sqrt{\kappa_i}$ are defined as the eigenbasis and the square roots of the eigenvalues of the reduced operator $\phi_S =  \tr_X\{\ket{\Phi}_{XS}\bra{\Phi} \}$. The Schmidt basis 
$\mathcal A =\{ \ket{\bar  i}_A ; i=0,...,d-1\}$ of system $A$ can be explicitly given by the orthonormal vectors $\ket{\bar  i} =\sum_x \sqrt{p(x)} \alpha_i^{(x)} \ket{x} / \sqrt{\kappa_i}$, where the $\alpha_i^{(x)} = \braket{i}{\varphi_x}$ are the coefficients of the signal states in the Schmidt basis $\mathcal B$. In the following, we omit the bar in $\ket{\bar i}$ for simplicity.

Furthermore, Alice's von Neumann measurement in the basis $\mathcal X$ on the larger space $\mathcal H_X$ is the Naimark extension of a measurement $\mathbf M_A = \{A_x \} $ with respect to rank-one POVM elements $A_x $ on the smaller space $\mathcal H_A$ given by
\begin{align}
A_x=p(x) \sqrt{\rho_A}^{-1}  \ketbra{\varphi_x^*}{\varphi_x^*}  \sqrt{\rho_A}^{-1} \label{Ax}.
\end{align}
Here we define the density matrix of Alice's reduced state as
\begin{align}
\rho_A = \tr_S\{\ket{\Phi}_{AS}\bra{\Phi} \}  \label{phiA},
\end{align}
and the states 
\begin{align}
\ket{\varphi_x^*} = \sum_i \ket i  \braket{\varphi_x}{i}   = \sum_i \ket i \; ({\alpha}_i^{(x)})^* \label{varphi_x*}  ,
\end{align}
where the symbol $*$ denotes the complex conjugate with respect to the Schmidt basis $\mathcal A$.
The operators $A_x$ are positive, sum up to the identity, and satisfy the property $\tr_A\{A_x \otimes \mathbbm 1 \ketbra{\Phi}{\Phi}\}=p(x) \ketbra{\varphi_x}{\varphi_x}$.

\begin{figure}
\includegraphics[width=0.5\textwidth]{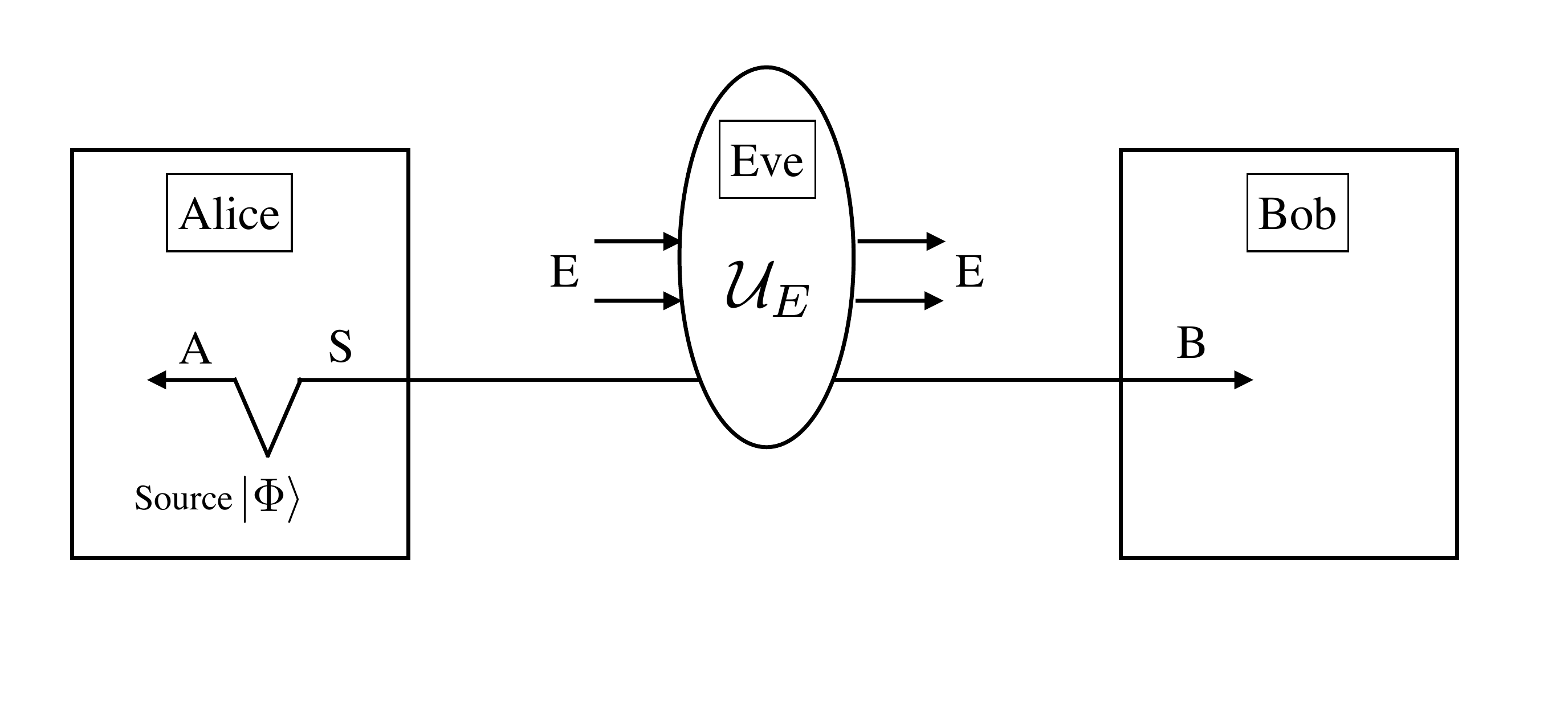}
\caption{\label{fig:setup} In the source-replacement scheme, Alice prepares the entangled state $\ket \Phi$. The system $A$ is kept by Alice, while the system $S$ is sent through the quantum channel to Bob.  Eve attaches ancillas to the signal states and performs a unitary transformation on the joint system $SE$, transforming it to $BE$. She resends the system $B$ to Bob. After Eve's interaction, Alice and Bob no longer share a perfect copy of $\ket{\Phi}$, but a bipartite state $\rho_{AB}$, which is only partially characterized by their observations. }
\end{figure}

After Eve's interaction with the signal states, but prior to Alice and Bob's measurements, the state held by Alice and Bob is described by an unknown (mixed) state $\rhoAB$ instead of a perfect copy of the source state. Nevertheless, Alice and Bob have some information about $\rhoAB$ due to their measurements. They can constrain the form of $\rhoAB$ from the probability distribution of their measurement outcomes 
\begin{equation}
p(x,y)=\tr \{ A_x \otimes B_y \; \rho_{AB} \} \label{pxy}
\end{equation}
in a process called parameter estimation.
In addition, Alice and Bob know that Alice's reduced density matrix of $\rho_{AB}$ remains unchanged, as already anticipated in equation (\ref{phiA}), because the system A never leaves Alice's lab (see Fig. \ref{fig:setup}). 
However, unless Alice and Bob's measurements are sufficient to obtain a tomographically complete parametrization of $\rhoAB$, there could be many states $\rhoAB$ that are compatible with $p(x,y)$ and $\rho_A$. 
For what follows, it is useful to make the following definition
\begin{definition} \label{def_Gamma}
The set $ \Gm$ contains all bipartite states $\rho_{AB}$ that are compatible with the measurement outcomes $p(x,y)$ and that have a given reduced state $\rho_A$.
\end{definition}

In the source-replacement scheme Eve attaches ancillas (defined on the system $E$) to the second half of the source state $\ket \Phi_{AS}$ followed by a unitary transformation $U_E$ which takes the composite system $SE$ to $BE$. She then keeps the transformed ancillas for herself, and resends the remaining system $B$ to Bob (see Fig. \ref{fig:setup}). The mixed state $\rhoAB$ is the result of Eve's interaction with the signal states. Eve's unitary transformation $U_E$ is equivalently characterized by the purification $\ket{\Psi}_{ABE}$ of $\rh{AB}$ on the dilated space $\mathcal H_{ABE}$, where the dimension of the purifying system $E$ is the same as the dimension of $AB$. In order to guarantee unconditional security of the protocol, we must assume that Eve can exploit everything allowed by quantum mechanics for her attack, which is realized by giving her full control over $\ket \Psi_{ABE}$. 

Note that to each $\rhoAB$ an entire class of purifications $\ket \Psi_{ABE}^{W} = \mathbbm  1_{AB} \otimes W_{E} \ket \Psi_{ABE}$ can be constructed, where $W$ is local unitary transformation on Eve's system. In what follows such local transformations on Eve's system are irrelevant.

\section{Key rate} \label{key_sec}

The security proof presented in Refs. \cite{devetak05a, renner05b, kraus05a} provides a bound on the rate at which Alice and Bob can extract a secret key. The proof is valid for collective attacks and for one-way classical communication. In many cases the proof can also be extended to hold for coherent attacks and two-way communication \cite{renner07a,renner05a}.

Given Alice, Bob and Eve share the purification $\ket \Psi$ of the state $\rhoAB$. After Alice and Bob measure their systems with respect to $A_x$ and $B_y$, they share the tripartite classical-classical-quantum (ccq) state \cite{devetak05a}
\begin{equation}
\rho_{XYE} =\sum_{x,y} p(x,y) \ketbra{x}{x}_X \otimes \ketbra{y}{y}_Y \otimes \rho_{E}^{xy} \label{rhoXYE}
\end{equation}
where $\ket{x}$ and $\ket {y}$ are two sets of orthonormal bases, and $\rho_{E}^{xy} = \tr_{AB} \{ A_x \otimes B_y \otimes \mathbbm 1_E \ketbra{\Psi}{\Psi}\} / p(x,y)$ are Eve's quantum states conditioned on the event that Alice and Bob's outcomes were $x$ and $y$.

Using error correction and privacy amplification, Alice and Bob extract a secret key from the ccq state $\rho_{XYE}$. A typical choice for the error correction is one-way reconciliation, in which the data of one party is set as a reference for the key, and the other party must correct her or his noisy data to match the reference. For our purposes, we will consider protocols with direct reconciliation, that is, Alice's data serves as the reference key and Bob corrects his data accordingly. 
The rate, established in \cite{devetak05a, kraus05a, renner05b}, at which an unconditionally secure key against collective attacks can be extracted is given by
\begin{equation}
r(\rho_{XYE})=  I(X:Y)-\chi(X:E)\label{r},
\end{equation}
where $I(X:Y)=H(X)+H(Y)-H(X,Y)$ is the classical mutual information of Alice and Bob's data, and $\chi (X:E) = H(X) + S(E) - S(X,E)$ is the Holevo quantity or quantum mutual information between Alice and Eve. $H$ and $S$ denote the Shannon entropy and the von Neumann entropy, respectively. The Holevo quantity is explicitly given by
\begin{eqnarray}
\chi (X:E) =S(\rho_{E})-\sum_{x} p(x) S(\rho_{E}^{x})  \label{Holevo}.
\end{eqnarray} 
where $\rho_{E}^{x}$ is Eve's state conditioned on Alice's value $x$ and $\rho_{E}=\sum_{x} p(x) \rho_{E}^{x} $ is Eve's reduced state.
For the practical calculation of the Holevo quantity, an explicit reference to the system $E$ can be eliminated, because the entropies $S(\rho_E)$ and $S(\rho_E^x)$ can be expressed in terms of quantities on the systems $AB$: If the state $\ket \Psi$ is pure, then $S(\rho_{AB}) = S(\rho_{E})$. If, furthermore, Alice uses rank-one POVM elements, then the conditional states $\rho_{BE}^x= \tr_A \{A_x \otimes \mathbbm 1_{BE} \ketbra{\Psi}{\Psi}\} / p(x)$ are pure, and therefore $S(\rho_{E}^x) = S(\rho_{B}^x)$. In this situation, the Holevo quantity simplifies to 
\begin{eqnarray}
\chi (X:E) =S(\rho_{AB})-\sum_{x} p(x) S(\rho_{B}^{x})  \label{HolevoAB}.
\end{eqnarray}

Usually, the key is not directly extracted from the state $\rho_{XYE}$, because the data $p(x,y)$ might be only weakly correlated. Alice and Bob typically postselect on highly correlated data before proceeding with the protocol. For example, Alice and Bob might ignore events which they measured in different bases - so called basis sifting - or they might discard data, where Bob did not record a detection event. 
Effectively, the key is extracted from a postselected state $\mathcal E(\rho_{AB}) $, which has again two classical registers $XY$ on Alice and Bob's side and a quantum register $E$ on Eve's side, but also additional classical registers carrying the information about the communication (announcements) between Alice and Bob that arise during the postselection.

Here we give a short description of the postselection that leads to $\mathcal E(\rho_{AB})$. For a more detailed formalism see appendix \ref{post_selection_app}: Alice and Bob announce some information about each signal to the public, that typically does not reveal any direct knowledge about the secret key. In the case of the sifting process, for example, they announce their basis choices. Depending on the announcement, the signal is either kept or discarded.
Let us denote the announcements of the kept signals by $u$. 
On the level of the quantum state $\rho_{AB}$, the measurement with respect to $\mathbf M_A$ and $\mathbf M_B$ followed by the announcement and the discarding is equivalently described by first a filtering operation on $\rho_{AB}$ followed by new measurements. 
The filtering yields transformed states $\rho_{AB}^u$ depending on the announcement $u$. The state held by Alice and Bob before the new measurement is then a convex combination of states $\rho_{AB}^u$ over the classical subsets $u$ with probability $p(u)$
\begin{equation}
\rho = \sum_u p(u) \rho_{AB}^u \otimes \ketbra{u}{u}.
\end{equation}
The new measurement is then performed on each $\rho_{AB}^u$ independently. 
In order to preserve the probability distributions of the measurement outcomes, we identify new pairs of POVMs $\mathbf M_A^u$ and $\mathbf M_B^u$ conditioned on the announcement $u$.
By giving to Eve the purification of each $\rho_{AB}^u$, we obtain new ccq states $\rho_{XYE}^u$ for each announcement $u$ after the measurement with respect to $\mathbf M_A^u$ and $\mathbf M_B^u$. The effective state $\mathcal E(\rho_{AB})$ after the post-selection is then described by the convex combination
\begin{equation}
\mathcal E(\rho_{AB}) = \sum_u p(u) \rho_{XYE}^u \otimes \ketbra{u}{u}.
\end{equation}

If we choose to extract the key from each $\rho_{XYE}^u$ independently, the effective key rate is given by
\begin{equation}
\bar r(\mathcal E(\rho_{AB})) = \sum_u p(u) r(\rho_{XYE}^u) \label{rE},
\end{equation}
where $r$ is the key rate given in equation (\ref{r}). Other choices of key extraction may combine different outcomes into one stream, but here we choose to analyze the simpler case of separate processing. 

By defining the total Holevo quantity and the total mutual information by 
\begin{align}
&\bar \chi(\mathcal E(\rho_{AB})) :=  \sum_{u} p(u) \chi_u(X:E) \label{ov_chi},\\
&\bar I(\mathcal E(\rho_{AB})) :=  \sum_{u}  p(u) I_u(X:Y) \label{ov_I},
\end{align}
where $\chi_u(X:E)$ and $I_u(X:Y)$ are the Holevo quantity and the mutual information of the state $\rho_{XYE}^u$, we can rewrite the effective key rate as
\begin{equation}
\bar r(\mathcal E(\rho_{AB})) = \bar I(\mathcal E(\rho_{AB})) - \bar \chi(\mathcal E(\rho_{AB}))  \label{ov_r}.
\end{equation}

If Alice and Bob knew Eve's attack strategy, the calculation of the key rate would be straightforward. However, all Alice and Bob know is that they share a state $\rhoAB$ from the set $\Gm$ in Def. \ref{def_Gamma}. Therefore, Eve has the freedom to chose any attack, as long as it creates a state $\rhoAB$ that is compatible with $\Gm$. Among all these possible attacks, the one that generates the lowest key rate 
\begin{eqnarray}
 r_{\mathrm{min}} &=& \inf_{\rho_{AB} \in \Gm} \bar r(\mathcal E(\rho_{AB})) \label{rmin}
\end{eqnarray}
is defined as the optimal attack. We call the pure state corresponding to the optimal attack 
$\ket \Psi^{\mathrm{opt}}$ with the reduced state $\rhoAB^{\mathrm{opt}}$. 
If Alice and Bob want to guarantee that their protocol is secure, they must assume that Eve performed the optimal attack. Hence, they cannot generate a secret key at a rate higher than $r_{\mathrm{min}}$ for the given protocol.


\section{Properties of the mutual information and the Holevo quantity} \label{theorems_sec}

In this section we give three properties of the classical mutual information and the Holevo quantity. 

At first, we introduce a new notation for the mutual information and the Holevo quantity, that is more convenient for our purposes throughout the rest of this paper. 
Let Alice and Bob share a quantum state $\rho_{AB}$, which they measure with respect to the POVM $\mathbf M_{AB} = \{ A_x \otimes B_y : A_x \in \mathbf M_A, B_y \in \mathbf M_B \}$. We always assume that Eve holds the purification $\ket \Psi$ of $\rho_{AB}$.
Instead of denoting $I(X:Y)$ with the dependence on the registers $XY$, we denote the mutual information by 
\begin{align}
I(\rho_{AB}, \mathbf M_{AB} ) := I(X:Y).
\end{align}
By specifying the quantum state $\rho_{AB}$ and the POVM $\mathbf M_{AB}$, the measured state on the registers $XY$ is entirely defined. 
On the other hand, the Holevo quantity can be directly calculated from the cq-state $\rho_{XE}$ that emerges after Alice's measurement of $\ket \Psi$ and after tracing over Bob's system. 
We write 
\begin{align}
\chi(\rho_{AB}, \mathbf M_A) := \chi(X:E)
\end{align}
for the Holevo quantity implying that there is a step from $\rho_{AB}$ to $\ket \Psi$.

In the first theorem of this section we show that the classical mutual information $I(\rho_{AB}, \mathbf M_{AB})$ is convex over $\rho_{AB}$ with fixed probability distribution $p(x) = \tr\{A_x \: \rho_{A}\}$. We call this feature ``weak convexity" to indicate that convexity only holds with the restriction on $p(x)$. 
\begin{theorem} \label{th_weakconvexity} \textbf{ (Weak convexity).}
Given the states $\rho_{AB}$, $\sigma_{AB}$ and the convex sum  $\bar \rho_{AB} = \lambda \rho_{AB} + (1-\lambda) \sigma_{AB}$ for $\lambda \in [0,1]$ with probability distributions $p(x,y) = \tr \{A_x \otimes B_y \; \rho_{AB} \}$, $q(x,y) = \tr \{A_x \otimes B_y \; \sigma_{AB} \}$ and $\bar p(x,y) = \lambda p(x,y) + (1-\lambda) q(x,y) $.
If the probability distributions satisfy $p(x) = q(x)$ for all $x$, then the mutual information is convex in the sense that
\begin{equation}
I(\bar \rho_{AB}, \mathbf M_{AB}) \leq \lambda I(\rho_{AB}, \mathbf M_{AB}) + (1-\lambda) I(\sigma_{AB}, \mathbf M_{AB}).
\end{equation}
\end{theorem}
The proof of this theorem is given in appendix \ref{proof_weakconvexity_app}. 

Second, we show that the Holevo quantity $\chi(\rho_{AB}, \mathbf M_A)$ is concave as a function of $\rho_{AB}$. 
\begin{theorem} \label{th_concavity} \textbf{ (Concavity).}
Given the states $\rho_{AB}$, $\sigma_{AB}$ and the convex sum $\bar \rho_{AB} = \lambda \rho_{AB} + (1-\lambda) \sigma_{AB}$ for $\lambda \in [0,1]$. Then, the Holevo is concave, meaning that it satisfies the property
\begin{equation}
\chi(\bar \rho_{AB},\mathbf M_A) \geq \lambda \chi(\rho_{AB}, \mathbf M_A) + (1-\lambda) \chi(\sigma_{AB},\mathbf M_A).
\end{equation}
\end{theorem}
The proof of this theorem is in appendix \ref{proof_concavity_app}.

Since we use postselection in our protocols, we need to extend these theorems to hold for the key rate $\bar r(\mathcal E(\rho_{AB}))$ in the following sense 
\begin{equation}
\bar r(\mathcal E(\bar \rho_{AB})) \leq \lambda \bar r(\mathcal E(\rho_{AB})) + (1-\lambda) \bar r(\mathcal E(\sigma_{AB})) \label{weakconvexr}.
\end{equation} 
This property (\ref{weakconvexr}) does not hold in general. For example, under certain postselection strategies, the restriction $p(x)=q(x)$ in theorem \ref{th_weakconvexity} may be violated. We will show later that for sifting on orthogonal basis states, the convexity property (\ref{weakconvexr}) holds.

Third, we show how $I(\rho_{AB}, \mathbf M_{AB})$ and $\chi(\rho_{AB}, \mathbf M_A)$ change under unitary transformations of the input state $\rho_{AB}$. 
\begin{lemma} \label{th_trafo}
Given the states $\rho_{AB}$ and $\sigma_{AB} = U \otimes V \rho_{AB} U^\dagger \otimes V^\dagger$. The mutual information and the Holevo quantity transform as follows:\begin{align}
&I(\sigma_{AB} , \mathbf M_{AB}) = I(\rho_{AB}, U^\dagger  \otimes V^\dagger \mathbf M_{AB} U \otimes V) \\
&\chi(\sigma_{AB} , \mathbf M_A) = \chi(\rho_{AB}, U^\dagger \mathbf M_A U )
\end{align}
where we define the sets 
\begin{align}
&U^\dagger \mathbf M_A U := \{ U^\dagger A_x U \} \label{UMU}\\
&V^\dagger \mathbf M_B V := \{ V^\dagger B_y V \} \label{UMUB}
\end{align}
and the set
\begin{align}
U^\dagger \otimes V^\dagger \; \mathbf M_{AB} \; U \otimes V 
:= \{ U^\dagger \otimes V^\dagger  ( A_x \otimes B_y) \;U \otimes V\}.
\end{align}
\end{lemma}
The proof of this lemma is based on the cyclic property of the trace, and that the von Neumann entropy is unitarily invariant: $S(U \rho U^\dagger) = S(\rho)$.

This lemma will prove useful in the next section, where we identify $U$ and $V$ with the unitary representations of the symmetry groups governing $A_x$ and $B_y$.


\section{Symmetries in protocols} \label{symmetries_sec}

In this section we introduce a scenario, in which the set $\Gm$ can be reduced to a set $\bar \Gm$, which contains only states with a certain symmetry corresponding to the symmetries of the signal states.

\subsection{Symmetries of the signal states and measurements}

Let G be a group with a unitary representation $\left \{ U_g ; g \in G \right \}$. A set of states $\mathbf S$ is $G$-invariant, if for all states $\ket{\varphi_x} \in \mathbf S$ and all $g \in G$ the state
\begin{equation}
\ketbra{\varphi_{g(x)}}{\varphi_{g(x)}}  := U_g \ketbra{\varphi_x}{\varphi_x} U_g^\dagger \label{Ginvariant}
\end{equation}
is also in $\mathbf S$. Here the index $g(x)$ denotes the index of the state $U_g \ket{\varphi_x}$. 
The following lemma describes the symmetry properties of the POVM elements $A_x$ and the reduced state $\rho_A$ in the source replacement picture (\ref{Ax}, \ref{phiA}):
\begin{lemma} \label{lemma_Gstarinv}
If the initial probability distribution $p(x)$ is uniform ($p(x) = 1/| \mathbf S| $ for all $x$) and the signal states are $G$-invariant, then the set of POVM elements $A_x$ and the reduced state $\rho_A$ are $G^*$-invariant, namely
\begin{eqnarray}
&U_g^* A_x U_g^T &= A_{g(x)} \label{Axsym}, \\
&U_g^* \rho_A U_g^T &= \rho_A \label{phiAsym}.
\end{eqnarray}
The symbols $*$ and $T$ denote the complex conjugate and the transpose with respect to the fixed Schmidt basis $\mathcal B$. 
\end{lemma}

We prove lemma \ref{lemma_Gstarinv} in appendix \ref{Ax_app_sec}.
Note that, with our particular definition of $*$ and $T$ with respect to the Schmidt basis, the operators $U_g^*$ and $U_g^T$ are well-defined.

In the following, we consider only protocols in which Bob's measurement operators $B_y$ are equipped with the $G$-invariance
\begin{eqnarray}
U_g B_y U_g^\dagger = B_{g(y)} \label{Bysym}.
\end{eqnarray}

\subsection{Parameter estimation with symmetries} \label{scenario_subs}

The symmetries in the signal states alone do not guarantee that the optimal eavesdropping attack is symmetric. Moreover, the observations and the post processing of the measured data also need to satisfy certain symmetry criteria. 

In many protocols only averaged measurement quantities are kept for the parameter estimation. For example, often only the quantum bit error rate (QBER) averaged over all the signal states is monitored.
In this section, and for the rest of this paper, we consider the scenario where Alice and Bob only keep averaged measurement quantities, and where the initial probability distribution $p(x)=1/|\mathbf S|$ of the signals is uniform.
In this scenario, Alice and Bob calculate a linear function $Q$ of the probability distribution $p(x,y)$ with the invariance property
\begin{equation}
Q[  p(x,y) ] = Q [  p(g(x), g(y)) ] \quad \forall g \in G \label{Qinvariance}.
\end{equation}
where the distribution
\begin{equation}
p(g(x),g(y)) = \tr \{ A_{g(x)} \otimes B_{g(x)} \rhoAB \}
\end{equation}
is generated by relabeling the POVM elements $A_x \otimes B_y$ by $A_{g(x)} \otimes B_{g(y)}$. Later on, we will identify $Q$ with the average error rate. 

From now on, Alice and Bob's knowledge about $\rho_{AB}$ is solely described by the average quantity $Q$ instead of the more detailed distribution $p(x,y)$.
In the previous section the set of all states that are compatible with the measurement data was the set $\Gm$ in Def. \ref{def_Gamma}.  Now, since the average quantity $Q$ is a coarse-grained version of $p(x,y)$, the set of states which are compatible with $Q$, $\Gm_{\mathrm{ave}}$, is a superset of $\Gm$, containing all states of the form
\begin{equation}
\rho_{AB}^{(U_g)}=U_g^* \otimes U_g \rho_{AB} (U_g^* \otimes U_g)^\dagger  \quad \forall g \in G \label{rhoUg},
\end{equation}
where $\rho_{AB} \in \Gm$.
The states $\rho_{AB}^{(U_g)}$ have the properties that 
(i) they are compatible with $Q$, and (ii) their reduced state $\tr_{B} \{ \rho_{AB}^{(U_g)} \} $ is equal to $\rho_A$. 

\begin{figure}
\includegraphics[width=0.5\textwidth]{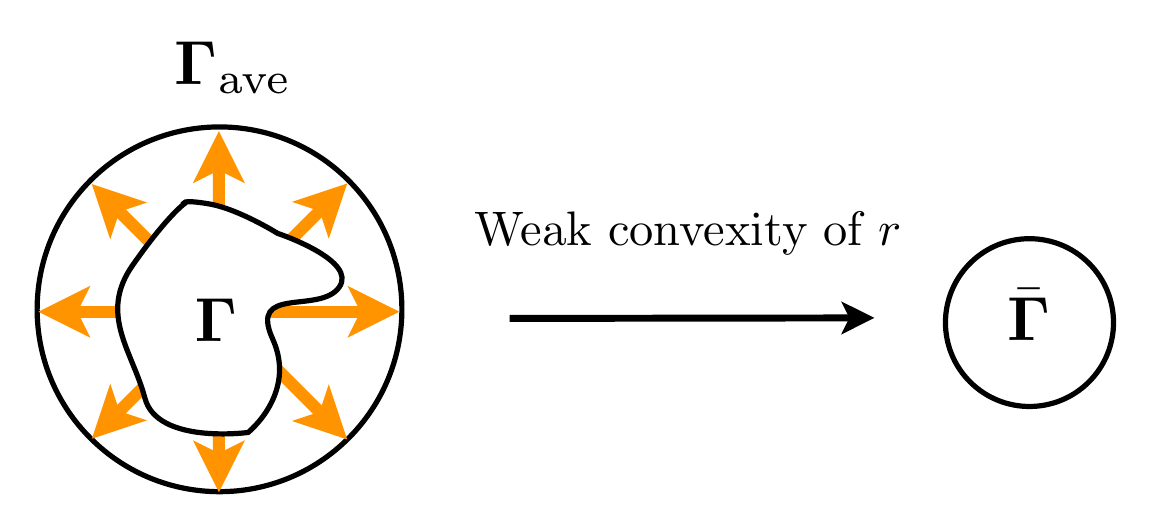}
\caption{\label{fig:Gamma_bar} By only using averaged measurement quantities $Q$ for parameter estimation, the set $\Gm$ is replaced by a bigger set $\Gm_{\mathrm{ave}}$. Using the weak convexity of the key rate, the optimal attack can be chosen from a symmetrized set $\bar \Gm$. }
\end{figure}

Let us now define a map that takes a bipartite state $\rho_{AB}$ and ``symmetrizes'' it with respect to the group $G$ by averaging over all the $\rho_{AB}^{(U_g)}$. In the literature this map is commonly known as twirling,
\begin{equation}
\mathcal T ^{G}[\rho_{AB}] \equiv \bar \rho_{AB} =\frac{1}{|G|}\sum_{g \in G} \rho_{AB}^{(U_g)}  \label{barrho},
\end{equation} 
where $|G|$ is the number of group elements in $G$.
Due to the linearity of $Q$,  $\Gm_{\mathrm{ave}}$ also contains all the states $\bar \rho_{AB}$. 

The twirling map
$\mathcal T^G$ maps the set  $\Gm_{\mathrm{ave}}$ to a subset, $\bar \Gm$, which contains only states of the form $\bar \rho_{AB}$. Each $\bar \rho_{AB}$ has the property that it commutes with all $U_g^* \otimes U_g$,
\begin{equation}
\left[\bar\rho_{AB}, U_g^* \otimes U_g\right]=0 \quad \forall g \in G \label{commutator}.
\end{equation}
The purification of a twirled state $\bar \rho_{AB}$ can be chosen to satisfies the following invariance:
\begin{equation}
U_g^* \otimes U_g \otimes U_g \otimes U_g^* \ket{\Psi}=\ket{ \Psi}\quad \forall g \in G \label{pure}.
\end{equation} 
The existence of this particular choice of the purification has been proven in Ref. \cite{christandl07a} for permutation groups, but the same proof holds for arbitrary groups as well.

\subsection{Symmetric attack}

One can restrict the search for the optimal attack $\rho_{AB}^{\mathrm{opt}}$ to a search over the set $\bar \Gm$, provided that the key rate of the particular protocol satisfies the convexity property,
\begin{equation}
\bar r(\mathcal E(\bar \rho_{AB})) \leq \frac{1}{|G|} \sum_{g \in G} \bar r(\mathcal E(\rho_{AB}^{(U_g)})) \label{PSconvexity},
\end{equation}
and the invariance property,
\begin{equation}
\bar r(\mathcal E(\rho_{AB})) = \bar r(\mathcal E(\rho_{AB}^{(U_g)})) \label{PSinvariance},
\end{equation}
under the corresponding symmetry group $G$. 
In this situation, $\rho_{AB}^{\mathrm{opt}}$ results from a symmetric attack and must lie in the subset $\bar \Gm \subset \Gm_{\mathrm{ave}}$ with the key rate given by
\begin{equation}
r_{\mathrm{min}} =  \inf_{\bar \rho_{AB} \in \bar \Gm} \bar r(\mathcal E(\bar \rho_{AB}))\label{rsym}.
\end{equation}

In Fig. \ref{fig:Gamma_bar} we represent schematically the transition from $\Gm$ to $\bar \Gm$. 
The symmetrized states $\bar \rho_{AB}$ are easily characterized using tools from representation theory. In appendix \ref{Schurs_lemma_app}, we show how to obtain $\bar \rho_{AB}$ using Schur's lemma.

In summary, in order to evaluate the key rate over the symmetrized set $\bar \Gm$, the protocol needs to exhibit sufficient symmetries, both in the quantum phase and in the classical phase: 
\begin{enumerate}
\item Symmetries in quantum phase:  the set of signal states $\mathbf S$ and Bob's POVM elements $B_y$ are $G$-invariant, and the a priori probability distribution $p(x)$ is uniform.
\item Coarse-grained parameter estimation: Alice and Bob restrict themselves to averaged measurement quantities $Q$ in the parameter estimation.
\item \label{cp2} Convexity and invariance of the key rate: the key rate $\bar r (\mathcal E(\rho_{AB}))$ satisfies
\begin{enumerate}
\item \label{cp2a} convexity $\bar r(\mathcal E(\bar \rho_{AB})) \leq \frac{1}{|G|} \sum_{g \in G}  \bar r(\mathcal E(\rho_{AB}^{(U_g)}))$,
\item \label{cp2b} invariance $\bar r(\mathcal E(\rho_{AB})) = \bar r(\mathcal E(\rho_{AB}^{(U_g)}))$.
\end{enumerate}
\end{enumerate}
In particular, the convexity and invariance properties depend strongly on the postselection procedure and must be checked for each protocol independently. 

\subsection{Examples of protocols with symmetric optimal attack: orthogonal bases as signal states}\label{ONB_sec}

Let us now construct a class of protocols where the set of signal states contains only complete sets of basis states, and where Alice and Bob postselect on events they measured in the same basis. This postselection is commonly referred to as sifting. We will show for this class of protocols that the convexity (\ref{PSconvexity}) and invariance (\ref{PSinvariance}) of the key rate always holds. Therefore, by choosing to keep only the average error rate $Q$ (defined below) for the parameter estimation, the optimal attack can always be assumed to be symmetric. 
 
First, we define the signal states and the measurements. 
Let us denote a basis of a $d$-dimensional Hilbert space by $\mathcal B_\beta = \{ \ket{\varphi_{(\beta,k)}} : k=0,...,d-1 \}$, where $\beta$ is the basis index. Note that, in the following the states $\ket{\varphi_{(\beta, k)}}$ carry two independent indices $(\beta, k)$ instead of only one.
The set of signal states of each protocol is then identified by
\begin{align}
\mathbf S_{\mathcal L} = \{\mathcal B_\beta: \beta \in \mathcal L \},
\end{align}
where $\mathcal L$ is the set from which the bases $\beta$ are drawn. For each protocol the set $\mathcal L$ is fixed and contains $|\mathcal L|$ elements. 
If each signal state $\ket{\varphi_{(\beta, k)}}$ is chosen with equal a priori probability $p(\beta, k) = 1/(d \cdot |\mathcal L|)$, Alice's reduced state $\rho_A$ in equation (\ref{phiA}) is proportional to the identity 
$ \rho_A = \mathbbm 1 / d$.
Therefore, Alice's POVM elements in equation (\ref{Ax}) reduce to the projectors  
\begin{equation}
A_{(\beta, k)} = \frac{1}{|\mathcal L|} \ketbra{\varphi_{(\beta, k)}^*}{\varphi_{(\beta, k)}^*}   \label{Axproj}
\end{equation}
with $\ket{\varphi_{(\beta, k)}^*} = \sum_i \ket i  \braket{\varphi_{(\beta, k)}}{i} $ defined in equation (\ref{varphi_x*}).
Furthermore, we construct Bob's POVMs to be isomorphic to Alice's:
\begin{align}
B_{(\beta,k)} =  \frac{1}{|\mathcal L|}  \ketbra{\varphi_{(\beta, k)}}{\varphi_{(\beta, k)}} \label{Byproj}.
\end{align}

Second, Alice and Bob postselect on those measurement outcomes, which they performed in the same basis. In this particular case, Alice and Bob's announcement $u$ is the basis $\beta$. 

We show three properties in appendix \ref{ONB_PS_app} for these protocols under post selection, which we will use to prove the convexity and the invariance of the effective key rate $\bar r(\mathcal E(\rho_{AB}))$: 
(i) The measurements conditioned on $u$ are simply renormalized versions of the original POVMs: 
\begin{align}
&\mathbf M_A^u = \{ |\mathcal L| \; A_{(\beta, k)} : \beta = u \} \label{MAu} \\
&\mathbf M_B^u = \{ |\mathcal L| \; B_{(\beta, k)} : \beta = u \}, \label{MBu}
\end{align}
(ii) the filtered states are independent of $u$ and satisfy $\rho_{AB}^u = \rho_{AB},$ and 
(iii) the probability distribution $p(u) = \frac{1}{|\mathcal L|}$ is uniform.
Additionally, since $\rho_A =\mathbbm 1/d$ and the $\mathbf M_A^u$ is a von Neumann measurement, the marginals $p_u(x) := \tr \{ \ketbra{\varphi_{(\beta, k)}^*}{\varphi_{(\beta, k)}^*} \rho_A \} = \frac{1}{d}$ are uniform. 

Using the uniform marginals $p_u(x)$ in combination with theorem \ref{th_weakconvexity}, each term $I(\rho_{AB}, \mathbf M_{AB}^u)$  is convex in $\rho_{AB}$. Moreover, due to the uniform distribution of $p(u)$, the convexity property immediately transfers to the convex sum
\begin{align}
\bar I(\mathcal E( \rho_{AB})) = \frac{1}{|\mathcal L|} \sum_u I(\rho_{AB}, \mathbf M_{AB}^u) \label{IONB}
\end{align}
Similarly, from theorem \ref{th_concavity} we conclude that each term $\chi(\bar \rho_{AB}, \mathbf M_A^u)$ is concave in $\rho_{AB}$, which transfers also to
\begin{align}
\bar \chi(\mathcal E( \rho_{AB}))= \frac{1}{|\mathcal L|} \sum_u \chi(\rho_{AB}, \mathbf M_{A}^u)\label{chiONB}
\end{align}
by the same argument.
The convexity of the effective key rate $\bar r(\mathcal E(\rho_{AB}))$ now follows immediately form the definition in equation (\ref{ov_r}). 

Next, we show the invariance of $\bar r(\mathcal E(\rho_{AB}))$ under the symmetry group $G$. Since any unitary acts like a basis transformation, the sets $\mathbf M_A^u$ and $\mathbf M_B^u$ effectively inherit the $G^*$- and $G$-invariance from the individual POVM elements $A_{(\beta,k)}$ and $B_{(\beta,k)}$. More precisely, the sets
\begin{align}
&\mathbf M_A^{g(u)} := U_g^* \mathbf M_A^u U_g^T, \label{GMAu}\\
&\mathbf M_B^{g(u)} := U_g \mathbf M_B^u U_g^\dagger \label{GMBu},
\end{align}
are again POVMs corresponding to the announcement with index $g(u)$ in the protocol. 

Using the definitions (\ref{GMAu}) and (\ref{GMBu}), and applying lemma \ref{th_trafo}, each component of $\bar r(\mathcal E(\rho_{AB}))$ transforms as follows under unitaries:
\begin{align}
&I(\rho_{AB}^{(U_g)} , \mathbf M_{AB}^{g(u)}) = I(\rho_{AB} , \mathbf M_{AB}^u)\\
&\chi(\rho_{AB}^{(U_g)} , \mathbf M_{A}^{g(u)}) = \chi(\rho_{AB} , \mathbf M_{A}^u).
\end{align}
Due to the uniform distribution of $p(u)$ and the $G^*$- and $G$-invariance of the POVMs, the invariance property of the convex sums $\bar I(\mathcal E( \rho_{AB}))$ and $\bar \chi(\mathcal E( \rho_{AB}))$ as well as $\bar r(\mathcal E(\rho_{AB}))$ follows directly.

In summary, if the average error rate $Q$ in definition \ref{error_rate_def} is used in the parameter estimation, the optimal attack lies in the symmetric subset $\bar \Gm$ for this class of protocols, and the key rate $r_{\mathrm{min}}$ can be calculated according to equation (\ref{rsym}) without loss of generality.

\begin{definition} \label{error_rate_def}
The average error rate is the probability that Alice sent the signal state $\ket{\varphi_{(\beta, k)}}$, but Bob received an orthogonal state $\ket{\varphi_{(\beta, k')}}$ ($k' \neq k$), averaged over all $k$ and all bases $\beta$
\begin{eqnarray}
&&Q=\frac{1}{|\mathcal L|} \sum_{\beta \in \mathcal L} Q^{\beta} \label{MUBQ},
\end{eqnarray}
where $Q^{\beta}$ is the average error rate found in each basis 
\begin{equation}
Q^{\beta} = \sum_{\substack{k, k' \\  k' \neq k}}  \tr \left \{ \ketbra{\varphi_{(\beta, k)}^*}{\varphi_{(\beta, k)}^*} \otimes \ketbra{\varphi_{(\beta, k')}}{\varphi_{(\beta, k')}} \rho_{AB} \right \} , \label{Qb}
\end{equation}
and $|\mathcal L|$ is the number of bases in the set $\mathcal L$. 
\end{definition}

The average Uhlmann fidelity of Bob's states with respect to Alice's signal state is defined as
\begin{align}
F_B = \frac{1}{|\mathcal L|} \sum_{\beta \in \mathcal L } \sum_k  \tr \left \{ \ketbra{\varphi_{(\beta, k)}^*}{\varphi_{(\beta, k)}^*} \otimes \ketbra{\varphi_{(\beta, k)}}{\varphi_{(\beta, k)}} \rho_{AB} \right \} , \label{F_B}
\end{align} 
With the definition of $Q$ above, $F_B$ and $Q$ are related by the simple relation $F_B=1-Q$.


\section{Quantum Cloners} \label{cloners_sec} 

A quantum cloner is a map that creates two copies of quantum states $\varphi_x=\ketbra{\varphi_x}{\varphi_x}$ drawn from a set $\mathbf S$. 
Let us define three isomorphic Hilbert spaces $\mathcal H_A$, $\mathcal H_B$ and $\mathcal H_C$ each with dimension $d$. A cloner $\mathcal C$ is a completely positive and trace preserving map 
$\mathcal C : \mathcal H_A \to \mathcal H_B \otimes \mathcal H_C$
that takes a state $\varphi_x \in \mathcal H_A$ to $\mathcal C (\varphi_x) \in \mathcal H_B \otimes \mathcal H_C$.
The quality of each copy $k$ ($k=B, C$) is determined by the single-clone Uhlmann fidelity $f_k(\varphi_x, \mathcal C(\varphi_x))$ of the copy with respect to the original state $\varphi_x$. If the states $\varphi_x$ are pure, the Uhlmann fidelity reads 
\begin{align}
f_B(\varphi_x, \mathcal C(\varphi_x)) = \tr \{\ketbra{\varphi_x}{\varphi_x}_B \otimes \mathbbm 1_C \cdot \mathcal C(\varphi_x) \} \\
f_C(\varphi_x, \mathcal C(\varphi_x)) = \tr \{\mathbbm 1_B \otimes \ketbra{\varphi_x}{\varphi_x}_C  \cdot \mathcal C(\varphi_x) \} 
\end{align}

Instead of $f_k(\varphi_x, C(\varphi_x))$, it is often assumed that only the average fidelity
\begin{equation}
F_k = \frac{1}{|\mathbf S|} \sum_{\varphi_x \in \mathbf S} f_k(\varphi_x, \mathcal C(\varphi_x)) \label{averageFk}
\end{equation}
is of interest. The cloner is called optimal if copy $C$ emerges with maximal average fidelity $(F_C)$, while the fidelity $F_B$ of the copy $B$ has a fixed value.

The cloning transformation can also be described using the Choi-Jamiolkowski isomorphism with a non-maximally entangled state. 
For this purpose, let $\mathcal C$ act on the second half of a source state $\ket \Phi$ defined in equation (\ref{source}).  
This relates $\mathcal C$ to a positive operator $\sigma_{ABC} \in \mathcal H_A \otimes \mathcal H_B\otimes \mathcal H_C $ via the rule
\begin{equation}
\sigma_{ABC}= (\mathbbm 1 \otimes \mathcal C) \ketbra{\Phi}{\Phi} \label{sigma}.
\end{equation}
The trace-preserving property of $\mathcal C$ translates to $\sigma_A=\tr_{BC} \sigma_{ABC} = \rho_A$, where $\rho_A$ is the reduced state of the source state $\ket \Phi$ given in equation (\ref{phiA}). From $\sigma_{ABC}$ the map $\mathcal C$ can be recovered through the reverse transformation realized by
\begin{equation}
\mathcal C(\varphi_x) = \frac{1}{p(x)}\tr_A[A_x  \otimes \mathbbm 1_B \otimes \mathbbm 1_C \cdot  \sigABC ].
\end{equation}
where the $A_x$ are the POVM elements defined in equation (\ref{Ax}). The reverse transformation effectively corresponds to preparing the states $\ket{\varphi_x}$ in the source-replacement scheme for the cloner. 

\subsection{Covariant cloners} \label{covariant_subs}

We now give a description of quantum cloners based on the work of Refs. \cite{chiribella05a, dariano01a}. 
Let the set of quantum states $\mathbf S$ is G-invariant. If the figure of merit is the average fidelity, then for every cloning map $\sigABC$, the ``rotated" maps 
\begin{equation}
U_g^* \otimes U_g \otimes U_g \sigma_{ABC} (U_g^* \otimes U_g \otimes U_g)^\dagger
\end{equation}
for $g \in G$ yield the same average fidelity $F_k$ as $\sigABC$. Furthermore, due to the linearity of the trace, there exists a covariant cloning map
\begin{equation}
\bar \sigma_{ABC}= \sum_{g \in G}  U_g^* \otimes U_g \otimes U_g \sigma_{ABC} (U_g^* \otimes U_g \otimes U_g)^\dagger  \label{sigave}
\end{equation}
with the same average fidelity $F_k$ as $\sigABC$. As a consequence, we can always choose the cloning map to be a covariant map, without loss of generality. 
The covariant state in equation (\ref{sigave}) satisfies the commutation relation
\begin{equation}
[\bar \sigma_{ABC}, U_g^* \otimes U_g \otimes U_g]=0 \label{covcom}.
\end{equation}

\subsection{Strong covariant cloner} \label{strong_subs}

In this section we describe a subset of cloning maps with a stronger symmetry than the covariant cloners, called strong covariant cloners.

The unitary realization $U_C$ of the map $\mathcal C$ can be uniquely described by the purification 
of $\bar \sigma_{ABC}$ on the extended Hilbert space $\mathcal H^{\otimes 6}$. This unitary realization requires three ancillas. Some cloners, however, can be realized using only one ancilla $D$, with $U_{C}$ now acting on a smaller space $
 H_A \otimes \mathcal H_B \otimes \mathcal H_C \otimes \mathcal H_D$. 
In \cite{chiribella05a}, these cloners are called strong covariant cloners and are defined as follows: 
\begin{definition}\label{def_strong_cov}
A cloner is called strong covariant if it has a purification $\ket{\Sigma}_{ABCD}$ on $\mathcal H_{ABCD}$ with the property \begin{equation}
 U_g^* \otimes U_g \otimes U_g \otimes U_g^* \ket{\Sigma}_{ABCD}=\ket{\Sigma}_{ABCD}  \; \forall g \in G \label{strongcovariance} .
\end{equation}
\end{definition}
It is easy to see that the strong covariant cloners are a subset of the covariant cloners: for every strong covariant cloner $\ket \Sigma$, tracing over the fourth system returns a covariant state $\bar \sigma_{ABC}$. 

Since the set of covariant cloning maps $\bar \sigma_{ABC}$ is a convex set and the fidelity is a linear functional of the cloning map, the cloning problem is a convex optimization problem \cite{chiribella05a}. The optimal cloner is an extremal point of the convex set, that means, a map that cannot be written as a convex sum of other maps in the set. We want to find the extremal map with the maximal clone fidelity. 
In \cite{chiribella05a}, two theorems about extremal maps are given:
\begin{theorem}\label{theorem_strong_cov_extremal}
(Chiribella et al. \cite{chiribella05a}). Let $U_g$ be an irreducible representation of the group $G$, and let $K =  \{\bar \sigma_{ABC}\}$ denote the set of covariant cloning maps with respect to $G$ according to equation (\ref{covcom}). Then, every cloning map $\bar \sigma_{ABC}$, which allows a strong covariant purification is an extremal point of the convex set $K$.
\end{theorem}
This theorem states that the strong covariant maps are a subset of the extremal maps. 
The converse - that the extremal maps are a subset of the strong covariant maps - is not true in general. 
The next theorem, however, describes a special case in which the set of extremal maps and the set of strong covariant maps coincide:
\begin{theorem}\label{theorem_strong_cov_Pauli}
(Chiribella et al. \cite{chiribella05a}).  If the set of states $\mathbf S$ to be cloned is G-invariant under the generalized Pauli group $\Pi_d$, then the set of strong covariant cloning maps is equal to the set of extremal maps. 
\end{theorem}
Theorem \ref{theorem_strong_cov_Pauli} allows to restrict the search of the optimal cloner to strong covariant maps based on the symmetries of the signal states alone. 
In section \ref{properties_sec}  we give the definition of the generalized Pauli group.


\section{Connection between optimal cloners and optimal attacks } \label{connection_sec}

We identify the optimal attack in QKD with an optimal cloner if Eve's interaction $U_E^{\mathrm{opt}}$ coincides with the optimal cloning transformation $U_C^{\mathrm{opt}}$, or, in terms of the purifications, if $\ket \Psi^{\mathrm{opt}} = \ket \Sigma^{\mathrm{opt}}$.
The optimal attack is always chosen from the set of purifications $\mathbf \Delta$ defined as follows:
\begin{definition}\label{def_Delta}
We define by $\mathbf \Delta$ the set that contains the purifications $\ket{ \Psi}$ of the symmetrized states in $\bar \Gm$. All states in $\mathbf \Delta$ satisfy the symmetry condition
$U_g^* \otimes U_g \otimes U_g \otimes U_g^* \ket{ \Psi}=\ket{ \Psi}$ for all $g \in G$, and they are compatible with the averaged quantity $Q$ and fixed $\rho_A$.
\end{definition}
We observe that all eavesdropping attacks represented by the set $\mathbf \Delta$ correspond to representations of strong covariant cloners. 
We can therefore make the following conclusion: 
\begin{observation} \label{theorem_strong_cov_attack_cloner}
The optimal cloner can only be the optimal attack, if it is strong covariant. Otherwise, one can already conclude that $\ket \Psi^{opt} \neq  \ket \Sigma^{opt}$.
\end{observation}

At this point, the strong covariance property alone does not uniquely determine if the optimal attack is an optimal cloner. Even if the optimal cloner is strong covariant, we can only conclude that the optimal attack is an optimal cloner if the set $\mathbf \Delta$ contains exactly one state. Otherwise, in order to compare the optimal attack with the optimal cloner, we must perform the optimization.

\subsection{Pauli-invariant signal states} \label{properties_sec}

As mentioned in theorem \ref{theorem_strong_cov_Pauli}, a sufficient requirement for a cloning map to allow a strong covariant realization is the Pauli-invariance of the set of states to be cloned. Hence, if the signal states of a QKD protocol are Pauli-invariant, the corresponding cloning attack on that protocol is certainly realized by a strong covariant cloner. Therefore, we will focus our attention on protocols with signal states that are Pauli-invariant. 

First, let us define the generalized Pauli group 
\begin{definition}\label{def_Pauli}
The generalized Pauli group $\Pi_d$ in $d$ dimensions has $d^2$ elements. The set of unitaries 
\begin{equation}
 U_{r,s}= \sum_{k=0}^{d-1} \omega^{ks} \ketbra{k+r}{k}, \quad \omega= \e^{2 \pi i /d} \label{Pauli},
\end{equation}
for $r,s=0,...,d-1$ form an irreducible unitary representation of $\Pi_d$ on a d-dimensional Hilbert space. The group has two generators
\begin{align}
Z := U_{0,1}, \quad X := U_{1,0}
\end{align}
which generate the entire group by the following relation:
\begin{align}
U_{r,s} = X^r Z^s \quad r,s = 0,...,d-1.
\end{align}
\end{definition}

A state $\rho_{AB}$ that commutes with all $U_{r,s}^* \otimes U_{r,s}$ is Bell-diagonal
\begin{equation}
\rho_{AB} = \sum_{r,s=0}^{d-1} u_{r,s} \ketbra{U_{r,s}}{U_{r,s}}  \label{Belldiag},
\end{equation} 
with eigenvalues $u_{r,s} \geq 0$  that satisfy $\sum_{r,s} u_{r,s} = 1$. The eigenvectors 
\begin{align}
\ket{U_{r,s}} = \frac{1}{\sqrt{d}}\sum_k \omega^{ks} \ket{k+r} \ket{k}
\end{align}
are called Bell states and form a maximally entangled basis of $\mathcal H^{\otimes 2}$. 
As shown in Refs. \cite{cerf02a, chiribella05a}, the general form of the purification of $\rho_{AB}$ is given by
$\ket \Psi = \sum_{r,s} \sqrt{u_{r,s}} \ket{U_{r,s}}  \ket{U_{r,d-s}}$.
Since $U_{r,s}$ is an irreducible representation, we can use Schur's lemma in appendix \ref{Schurs_lemma_app} to characterize the reduced state $\rho_{A}= \frac{\mathbbm 1}{d}$, which is proportional to the identity.


\section{Examples with mutually unbiased bases} \label{examples_sec}

Mutually unbiased bases (MUBs), which were first introduced in Refs. \cite{ivanovic97a, ivanovic81a}, are a common choice for the signal states of QKD protocols. For example, 
in the qubit space, the BB84 and the 6-state protocol use 2 and 3 MUBs, respectively. In higher-dimensional Hilbert spaces, MUB protocols have been studied in Refs. \cite{bechmann00a, bruss02a, cerf02a, durt03a, durt04a, caruso05a}. 

MUBs are orthonormal bases $\mathcal B_\alpha =  \{\ket{\psi_1^\alpha}, \ket{\psi_2^\alpha}, ... , \ket{\psi_{d-1}^\alpha}\} $ on d-dimensional Hilbert spaces with the property $|\braket{\psi_k^\alpha}{\psi_{k'}^{\alpha'}} | = \frac{1}{\sqrt{d}}$ for all $k, k' = 0,...,d-1$ and $\alpha \neq \alpha'$.
In Ref. \cite{wootters89a}, it was shown that when $d$ is a prime number, there exist exactly $d+1$ MUBs. The eigenbases of the generalized Pauli operators $Z$ and $XZ^\beta$ for $\beta = 0,...,d-1$ form such MUBs \cite{bandy08a}. The eigenbasis of the operator $Z$ is denoted by the standard basis with the index $Z$,
\begin{align}
&\mathcal B_Z = \{\ket{\psi_1^{Z}}, \ket{\psi_2^{Z}}, ..., \ket{\psi_{d-1}^{Z}}\} \label{standard}, \\
&\ket{\psi_k^{Z}} = \ket k,
\end{align}
and the eigenbases of the operators $XZ^\beta$ with indices $\beta=0,...,d-1$ by
\begin{align}
&\mathcal B_\beta = \{\ket{\psi_1^{\beta}}, \ket{\psi_2^{\beta}}, ..., \ket{\psi_{d-1}^{\beta}}\} \label{Bbeta}, \\
&\ket{\psi_k^{\beta}} = \frac{1}{\sqrt d} \sum_{j=0}^{d-1} \omega^{-kj} \omega^{-\beta s_j} \ket{j}, 
\end{align}
where $s_j = \frac{1}{2}(d-j)(d+j-1)$, $\omega= e^{2 \pi i/d}$, and where $\ket j$ are the basis vectors of the standard basis $\mathcal B_Z$. 

Let us now consider protocols where the set of signal states $\mathbf S_{\mathcal L}$ contains any subset of these $d+1$ eigenbases.
In a slight generalization of the result of theorem 2.2 in \cite{bandy08a},
we can show that the action fo any Pauli operator $U_{r,s}$ on the eigenstates of the Pauli eigenbasis $\mathcal B_\alpha$ for $\alpha \in \{Z, 0, 1, ..., d-1 \}$ permutes the eigenstates without changing the basis index $\alpha$. Using this invariance, the set of signal states $\mathbf S_{\mathcal L}$ of any such protocol are also Pauli-invariant. 

Unfortunately, the full symmetry groups of the sets $\mathbf S_{\mathcal L}$ are not known explicitly. Thus, one cannot simply write down the general form of the symmetrized states $\bar \rho_{AB} \in \bar \Gm$.
However, in a first step, we can exploit the invariance of the set $\mathbf S_{\mathcal L}$ with respect to the generalized Pauli group. This partial symmetry implies that the optimal attack $\rho_{AB}^{\mathrm{opt}}$ must lie in the subset $\Gm_{\mathrm{Bell}}$ containing only Bell-diagonal states defined in equation (\ref{Belldiag}). 

Let us, therefore, calculate the key rate (\ref{ov_r}) for MUB protocols with Bell-diagonal states $\rho_{AB}$. For this purpose, we require the eigenvalue spectrum of each conditional state on Bob's side $\rho_B^{(\alpha,k)}$ for $\alpha \in \{Z, 0, 1, ..., d-1 \}$, given that Alice obtained the measurement outcome corresponding to the state $\ket{\psi_k^\alpha}$. We project Alice's system of the Bell state onto $\ket{\psi^{*\alpha}_k}$ and divide the result by a normalization constant $N$: 
\begin{align}
\rho_B^{(\alpha,k)} = \sum_{r,s} u_{r,s} \braket{\psi^{*\alpha }_k}{U_{r,s}} \braket{U_{r,s}}{\psi^{*\alpha }_k} / N \label{rhoBraw}.
\end{align}
In the following, all operations are modulo $d$, and in particular, the indices are to be understood modulo $d$. The overlaps in (\ref{rhoBraw}) are found to be
\begin{align}
&\braket{\psi^{*Z}_k}{U_{r,s}} = \frac{1}{\sqrt d} \omega^{(k-r)s} \ket{k-r}\\
&\braket{\psi^{* \beta}_k}{U_{r,s}} = \frac{1}{\sqrt d} \omega^{-kr - \frac{\beta}{2} (r-r^2)} \ket{\psi^\beta_{k-(s+\beta r)}} \label{overlapbeta}. 
\end{align}
for $Z$ and for $\beta \in \{0,...,d-1 \}$.
After reinserting the overlaps into (\ref{rhoBraw}) and using by $N= \frac{1}{d}$, we do an index substitution $y = s + \beta r$ in (\ref{overlapbeta}) to obtain the following expressions for the conditional states for $\alpha \in \{Z, 0, 1, ..., d-1 \}$
\begin{align}
&\rho_B^{(\alpha,k)} = \sum_y \lambda_y^\alpha \ketbra{\psi^\alpha_{k-y}}{\psi^\alpha_{k-y}}.
\end{align}
The set of eigenvalues $\mathbf \Lambda^\alpha = \{ \lambda_0^{\alpha}  , \lambda_1^{\alpha}  ,..., \lambda_{d-1}^{\alpha} \}$ for each $\rho_B^{(\alpha,k)}$ is independent of the index $k$ with the specific values
\begin{align}
&\lambda_y^{Z} =  \sum_{r=0}^{d-1} u_{y,r}    &&\mathrm{for }\;   Z \label{lambda_d} \\
&\lambda_y^{\beta} = \sum_{r=0}^{d-1} u_{r,y-\beta r}  &&\mathrm{for } \; \beta \in \{ 0,...,d-1\} \label{lambda_beta}.
\end{align}
The average error rate, the mutual information and the Holevo quantity (\ref{MUBQ}, \ref{IONB}, \ref{chiONB}) can now be calculated using the eigenvalue spectrum 
\begin{eqnarray}
&&Q=1- \frac{1}{|\mathcal L|} \sum_{\alpha \in \mathcal L} \lambda_0^\alpha \label{Qsym},\\
&&\bar I(\mathcal E(\rho_{AB})) = \log_2 d  - \frac{1}{|\mathcal L|} \sum_{\alpha \in \mathcal L} H(\mathbf \Lambda^\alpha), \\
&&\bar \chi(\mathcal E(\rho_{AB})) = S(\rho_{AB}) - \frac{1}{|\mathcal L|} \sum_{\alpha \in \mathcal L} H(\mathbf \Lambda^\alpha) \label{chilambda},
\end{eqnarray}
with the Shannon entropy $H(\mathbf \Lambda^\alpha) = -\sum_y \lambda_y^\alpha \log(\lambda_y^\alpha)$ and the von Neumann entropy $S(\rho) = -\tr \rho \log \rho$. 
The key rate (\ref{ov_r}) follows straightforwardly, 
\begin{equation}
\bar r(\mathcal E(\rho_{AB})) = \log d - S(\rho_{AB}) \label{rpropS}.
\end{equation} 

For the special protocols with $2$, $d$ and $d+1$ MUBs, we further confine our search for the optimal attack to smaller subsets $\tilde \Gm \subset \Gm_{\mathrm{Bell}}$. The procedure is essentially the same as the one we used in section \ref{symmetries_sec} to reduce to the subset $\bar \Gm$ from the set $\Gm_{\mathrm{ave}}$. For each state $\rho_{AB}$ in $\Gm_{\mathrm{Bell}}$, we generate an equivalence class of states $\{ \rho_{AB}^{(P_i)} ; i = 1 ,...,n \}$  by applying permutations $P_i$ to the eigenvalues of $\rho_{AB}$.  In contrast to the states $\rho_{AB}^{(U_g)}$ in equation (\ref{rhoUg}), the states $\rho_{AB}^{(P_i)}$ are not generated using the symmetry group of the signal states. However, since the key rate (\ref{rpropS}) is proportional to $S(\rho_{AB})$, all permuted states satisfy the invariance property $\bar r(\mathcal E(\rho_{AB})) = \bar r(\mathcal E(\rho_{AB}^{(P_i)}))$. Furthermore, we chose the permutations in such a way, that the $\rho_{AB}^{(P_i)}$ give the same error rate as $\rho_{AB}$. This ensures that the $\rho_{AB}^{(P_i)}$ are again in $\Gm_{\mathrm{Bell}}$. 
Consequently, since the convexity property (\ref{PSconvexity}) of the key rate holds for protocols with MUBs, we can conclude that the optimal attack is found in the subset $ \tilde \Gm \subset \Gm_{\mathrm{Bell}}$ containing only convex combinations,
\begin{align*}
\tilde \rho_{AB} =\frac{1}{n} \sum_i \rho_{AB}^{(P_i)} \in \tilde \Gm.
\end{align*}
Note that it suffices to check that the states $\rho_{AB}^{(P_i)}$ have the same average error rate $Q$ as $\rho_{AB}$ in order to be in the set $\Gm_{\mathrm{Bell}}$. We do not need to monitor the condition on $\rho_A$, because $\rho_A = \mathbbm 1 / d$ is automatically satisfied for any Bell-diagonal state. 

\subsection{$2$ MUBs}

We describe here how to obtain the states $\tilde \rho_{AB} \in \tilde \Gm$ for the example of the $2$ MUBs protocol with the signal states  $\mathbf S_{\mathcal L} = \{ \mathcal B_0, \mathcal B_Z \}$: given a Bell-diagnoal states $\rho_{AB}$. As mentioned above, we generate the (Bell-diagonal) permuted states $\rho_{AB}^{(P_i)}$ by keeping the error rate (\ref{Qsym}) of $\rho_{AB}$ invariant
\begin{align}
Q =  1 - \frac{1}{2} \left ( \lambda_0^0 + \lambda_0^Z \right ) = 1 -  \frac{1}{2} \left ( 2 u_{0,0} + \sum_{r=1}^{d-1} (u_{r,0} + u_{0,r}) \right )\label{Q2d}.
\end{align}
The invariance of $Q$ is guaranteed if the permutations $P_i$ leave the sets 
$\mathbf U_a = \{u_{0,0}\}$, $\mathbf U_b = \{u_{0,r}, u_{r,0} ; r=1,...,d-1\}$ and $\mathbf U_c= \{u_{r,s} ; r,s=1,...,d-1\}$ invariant.
Such permutations $P_i$ are, for example, independent permutations of the eigenvalues in each set. Therefore, in the convex combination $\tilde \rho_{AB}$, the average over all eigenvalues in each set will appear. In this particular case, $\tilde \rho_{AB}$ has three different types of independent eigenvalues $a$, $b$ and $c$ corresponding to the three sets $\mathbf U_a$, $\mathbf U_b$ and $\mathbf U_c$,
\begin{eqnarray}
&\tilde \rho_{AB} &= a \ketbra{U_{0,0}}{U_{0,0}} + c \sum_{ r,s = 1 }^{d-1} \ketbra{U_{r,s}}{U_{r,s}} \label{eq_2MUBs} +\\
&& b \sum_{ r =1 }^{d-1} \left( \ketbra{U_{r,0}}{U_{r,0}} +  \ket{U_{0,r}} \ket{U_{0,r}}\right) \nonumber.
\end{eqnarray}

\begin{figure*}
\includegraphics[scale=0.8]{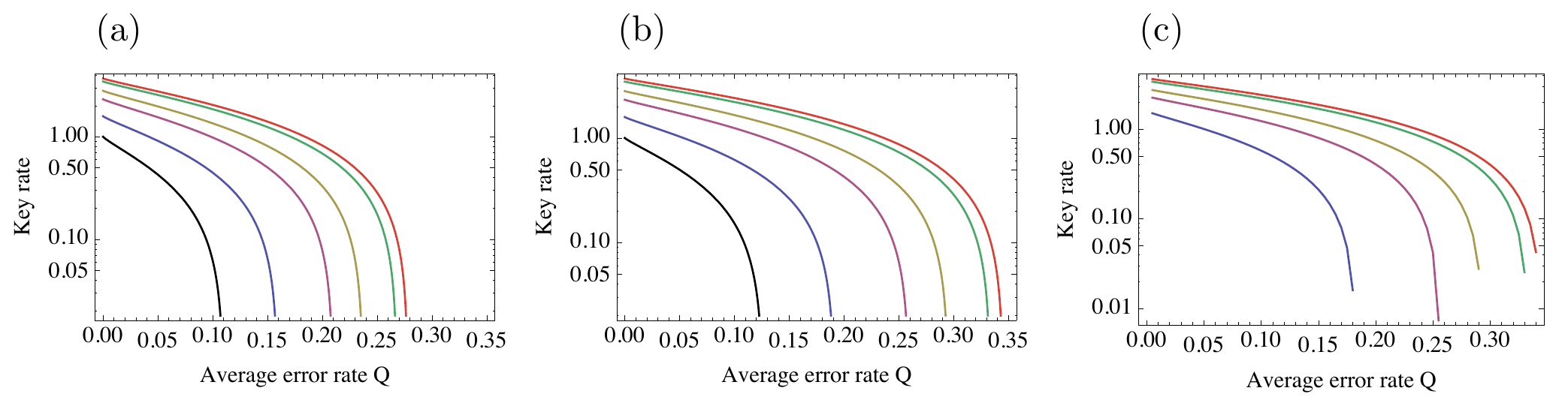}
\caption{\label{fig:key_rates}Key rates of protocols with (a) $d+1$, (b) $2$ and (c) $d$ MUBs for $d=2$ (black), $d=3$ (blue), $d=5$ (purple) , $d=7$ (yellow) , $d=11$ (green) and  $d=13$ (red). These plots do not serve as a comparison of the performance of the different protocols.  }
\end{figure*}

Now we can use the average error rate condition $Q = (d-1)b + (d-1)^2 c$ and the normalization condition $a+ 2(d-1)b +d^2 c =1$, to further reduce the number of independent eigenvalues to only one. Through an (analytical) optimization of the key rate over the free eigenvalue, we find the following values for $a$, $b$ and $c$ that describe the optimal attack
\begin{eqnarray}
a=(1-Q)^2, \quad b= \frac{Q(1-Q)}{d-1}, \quad c= \frac{Q^2}{(d-1)} \label{b2}.
\end{eqnarray}
These are exactly the same that also describe the optimal phase-covariant cloner in $d$ dimensions \cite{cerf02a}. The connection between optimal cloning and the optimal attack for $2$ MUBs was already  
conjectured in \cite{cerf02a}. 
The key rates for these protocols, which were independently obtained in Ref. \cite{sheridan10a},  are given by 
\begin{eqnarray}
r_{\mathrm{min}} = \log d +  2 (1-Q) \log(1-Q) + 2Q \log \left ( \frac{Q}{d-1} \right ) \nonumber.
\end{eqnarray}
They are plotted for $d=2,3,5,7,11,13$ in Fig. \ref{fig:key_rates}. Note that this plot is not intended to compare the performances of the different protocols. For a fair comparison, one must specify the channel model for which the key rates are drawn.

\subsection{$d+1$ MUBs} \label{d1MUBs_sec}

Consider protocols with the signal states $\mathbf S_{\mathcal L} = \{\mathcal B_Z, \mathcal B_0, \mathcal B_1 , ... \mathcal B_{d-1} \}$. In contrast to the case with 2 MUBs, these protocols are  tomographically complete. With the same strategy as for $2$ MUBs, we construct the set $\tilde \Gm$ for this situation: The error rate (\ref{Qsym}) of a Bell-diagonal state in this scenario is given by
\begin{align}
Q =  1 - \frac{1}{d+1} \left ( (d+1) u_{0,0} + \sum_{(r,s) \neq (0,0)} u_{r,s} \right )
\end{align}
where we used the relation that for $r \neq 0$, the sum $\sum_{\beta = 0}^{d-1} u_{r,-\beta r} = \sum_{\gamma = 0}^{d-1} u_{r,\gamma}$. The error rate defines the sets $\mathbf U_a = \{u_{0,0}\}$ with one eigenvalue, and $\mathbf U_{b} = \{ u_{r,s}; (r,s) \neq (0,0)\}$ with the remaining $d^2-1$ eigenvalues. Again, $\mathbf U_a$ and $\mathbf U_b$ determine the form of the states in the set $\tilde \Gm$, after averaging over all permutations $P_i$
\begin{eqnarray}
&\tilde \rho_{AB} &= a \ketbra{U_{0,0}}{U_{0,0}} + 
b \sum_{ (r,s) \neq (0,0)} \ketbra{U_{r,s}}{U_{r,s}} \label{eq_d1MUBs}.
\end{eqnarray}
In this situation, the average eigenvalues $a$ and $b$ are uniquely defined by the error rate $Q=d(d-1)b$ and the trace condition $a+ (d^2-1)b =1$
\begin{eqnarray}
a = 1 - \frac{d+1}{d}Q, \quad b = \frac{ Q}{d(d-1)} \label{bd1}.
\end{eqnarray}
As there is only one state in $\tilde \Gm$ for this protocol, the optimization of the key rate becomes trivial, and we can conclude that
the optimal cloner and the optimal attack are equal. 
This connection was already conjectured in \cite{cerf02a}. The cloner in this case is the optimal universal cloner in $d$ dimensions \cite{cerf00a, cerf00c}. For $d=2$, we recover the the 6-state protocol, where the optimal cloner is the universal cloner \cite{buzek96a, gisin97a}, which clones all the states on the Bloch sphere equally well. 
The key rates of these protocols are given by
\begin{eqnarray}
r_{\mathrm{min}} &=&  \log d + \frac{d+1}{d} Q \log \left ( \frac{Q}{d(d-1)} \right ) \\
&+&   \left (1 - \frac{d+1}{d}Q \right ) \log \left (1 - \frac{d+1}{d}Q \right )  \nonumber,
\end{eqnarray}
as independently shown in Ref. \cite{sheridan10a}. We plot the key rates for $d=2,3,5,7,11,13$ in Fig. \ref{fig:key_rates}. Again, the plot is not intended to compare the performance of the protocols. 

\subsection{$d$ MUBs}

The signal states of protocols using $d$ MUBs are $\mathbf S_{\mathcal L} = \{\mathcal B_0, \mathcal B_1 , ... \mathcal B_{d-1} \}$. Unlike the protocols with $d$ MUBs, the protocols analyzed here are not tomographically complete. The error rate
\begin{align}
Q = 1 - \frac{1}{d} \left ( d \; u_{0,0} + \sum_{r=0}^{d-1} \sum_{s=1}^{d-1} u_{r,s} \right )
\end{align}
defines three sets $\mathbf U_a = \{u_{0,0} \}$, $\mathbf U_b = \{u_{r,s} ; r=0,...,d-1, s=1,...,d-1\}  $ and $\mathbf U_c=\{u_{0,s} ; s=1,...,d-1\}$, which determine the form of $\tilde \rho_{AB} \in \tilde \Gm$
\begin{eqnarray}
&\tilde \rho_{AB} &= a \ketbra{U_{0,0}}{U_{0,0}} + c \sum_{ s = 1 }^{d-1} \ketbra{U_{0,s}}{U_{0,s}}  \label{eq_dMUBs}+\\
&& b \sum_{ r =0 }^{d-1} \sum_{s=1}^{d-1} \ketbra{U_{r,s}}{U_{r,s}} \nonumber .
\end{eqnarray}
The eigenvalues $a$, $b$ and $c$ are further constriced by the normalization condition $a+ d(d-1)b + (d-1)c =1$, and the error rate condition for $\tilde \rho_{AB}$
$Q = (d-1)^2 Êb + (d-1)c$.
We can express two of the three eigenvalues by
\begin{eqnarray}
a =  1+c-\frac{dQ}{d-1} \quad b= \frac{Q-(d-1)c}{(d-1)^2} . 
\end{eqnarray}
We perform an optimization of the key rate over the free eigenvalue $c$, and plot the numerically obtained key rates $r_{\mathrm{min}}$ in Fig. \ref{fig:key_rates} for different dimensions.   

\begin{figure}
\includegraphics[scale=1]{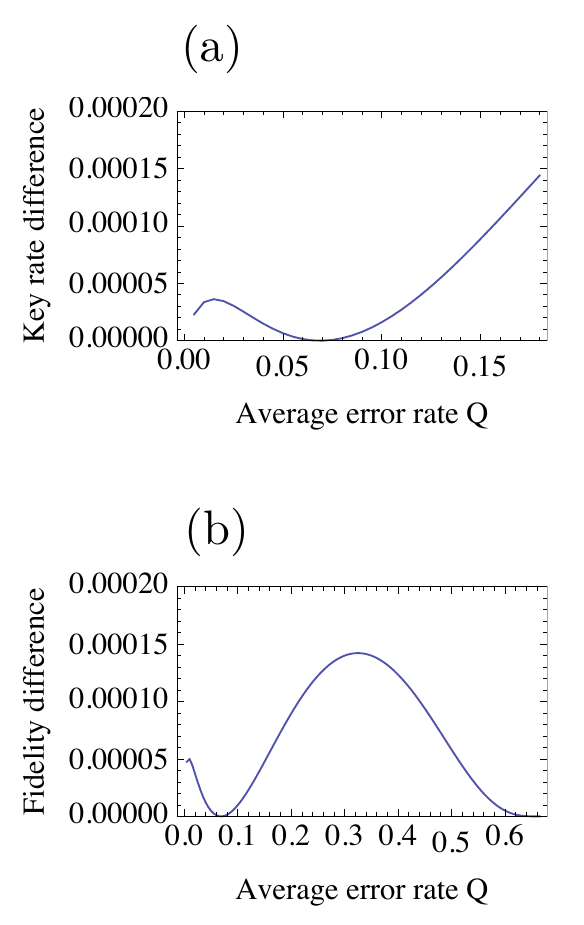}
\caption{\label{fig:difference} (a) A plot of the difference of the key rates $\delta r =  r_{\mathrm{MPC}} - r_{\mathrm{min}} $ for the scenarios where Eve uses the optimal cloner ($r_{\mathrm{MPC}}$),  and where she uses the optimal attack ($r_{\mathrm{min}}$) for $d=3$. (b) A plot of the difference of the fidelities $\delta F_E = F_E^{\mathrm{MPC}} - F_E^{\mathrm{attack}} $ for the scenarios where Eve uses the optimal MPC cloner ($F_E^{\mathrm{MPC}}$) and where she uses the optimal attack ($F_E^{\mathrm{attack}}$) for $d=3$.  }
\end{figure}

We compare the optimal attack to the optimal multiple phase-covariant (MPC) cloner $U_{\mathrm{MPC}}$ given in Ref. \cite{lamoureux05a}. This cloner copies all states of the form $\ket{\psi} = \frac{1}{\sqrt{d}} \sum_{j=0}^{d-1} e^{i \phi_j} \ket{j}$ for $\phi_j \in [0, 2\pi )$ optimally. 
If it is known that the eavesdropper performed an attack based on the optimal MPC cloner, Alice and Bob can expect some key rate $r_{\mathrm{MPC}}$.  
Numerical optimizations for $d=3, 5, 7, 11$ and $13$ show that $r_{\mathrm{MPC}}$ is always bigger (or equal) than the key rate $r_{\mathrm{min}}$. Therefore, the optimal MPC cloner is not the optimal attack: $U_{\mathrm{MPC}} \neq U_{E}^\mathrm{opt}$.    
In Fig. \ref{fig:difference} we plot the difference between the key rates $r_{\mathrm{MPC}} - r_{\mathrm{min}} $, and the difference between the fidelities $F_E^{\mathrm{MPC}} - F_E^{\mathrm{attack}}$  for $d=3$, where $F_E^{\mathrm{attack}}$ is Eve's fidelity when using the optimal attack.

We would like to remark that the $U_{\mathrm{MPC}}$ produces optimal copies of more than just the necessary $d$ MUBs. It is possible that there exists a cloner $U_d$ that provides copies of the $d$ MUBs with a higher fidelity $F_E^{d}$ (for fixed error rate) than $U_{\mathrm{MPC}}$: $F_E^{d} \geq F_E^{\mathrm{MPC}}$.
This raises the question, if the cloner $U_d$ could be the optimal attack $U_{E}^\mathrm{opt}$. 
To answer this, we turn the question around and ask from the cloning point of view, whether the optimal attack $U_{E}^\mathrm{opt}$ can play the role of the optimal cloner $U_d$. 
For this purpose, we compare Eve's fidelity $F_E^{\mathrm{attack}}$, when she used the optimal attack to the fidelity $F_E^{d}$, when she used $U_d$.
We know from our numerical optimization how $F_E^{\mathrm{attack}}$ compares to $F_E^{\mathrm{MPC}}$: we plotted the difference $F_E^{\mathrm{MPC}} - F_E^{\mathrm{attack}}$ in  Fig. \ref{fig:difference}. From this plot we see that, in general, $F_E^{\mathrm{MPC}} > F_E^{\mathrm{attack}}$. However, we know by construction of $U_d$ that the fidelity $F_E^{d} \geq F_E^{\mathrm{MPC}} $. By transitivity it follows that $F_E^d > F_E^{\mathrm{attack}}$, which proves that the optimal attack is not equivalent to the optimal cloner $U_d$ either. 

In table \ref{tab:summary} we summarize the results obtained for the MUB protocols. 
It turns out that in general the intuition that the optimal attack is always an optimal cloner proves wrong, as can be seen for the protocols with $d$ MUBs.

\begin{table*}
\caption{\label{tab:summary}Summary of optimal attack of protocols using MUBs in $d$ dimensions. For the protocols where the optimal attack is an optimal cloner we put a checkmark ($\checkmark$). Where the optimal attack is not an optimal cloner we put a cross ($\times$). The numerical optimizations cover the cases of $d$ MUBs up to $d=13$. }
\begin{ruledtabular}
\begin{tabular}{ccccccccc}
&&&& Number of MUBs  &&&&\\
Dimension & 2 MUBs & 3 MUBs & 4 MUBs & 5 MUBs &6 MUBs & $\hdots$ 
&d MUBs & d+1 MUBs \\ \hline
 2 &  $\checkmark$  & $\checkmark$&  & & & &  & \\
 3 & $\checkmark$  & $\times$  & $\checkmark$  & & & & & \\
 5 & $\checkmark$  &  &   & $\times$ & $\checkmark$ &  & & \\
 $\vdots$ & $\vdots$  && &  & &  & $\ddots$ &   \\
 d & $\checkmark$  &   &   &   & & & & $\checkmark$ \\
\end{tabular}
\end{ruledtabular}
\end{table*}


\section{Classes of protocols with the same optimal attack} \label{classes_sec}

We observe that for certain protocols with different signal states, the same attack $\rho_{AB}^{\mathrm{opt}}$ is found to be optimal. 
Given two protocols  $P$ and $P'$ with sets of signal states $\mathbf S$ and $\mathbf S'$ that are $G$- and $G'$-invariant, respectively. We consider only protocols where the signal states are orthogonal bases, and where Alice and Bob postselect on events in the same basis. These protocols were already described in section \ref{ONB_sec}. We denote the average error rate for each protocol by $Q$ and $Q'$. 
Let us denote Alice and Bob's POVM elements by $A_x$ and $B_y$ for the protocol $P$  and by $A'_{x'}$ and $B'_{y'}$ for the protocol $P'$ which are given in equation (\ref{Axproj}) and (\ref{Byproj}). The POVMs conditioned on the basis announcement $u$ given in equations (\ref{MAu}) and (\ref{MBu}) are denoted by $\mathbf M_A^u$ and $\mathbf M_B^u$ for protocol $P$ and by $\mathbf {M'}_A^{u'}$ and $\mathbf {M'}_B^{u'}$ for protocol $P'$. 
The sets characterizing the possible symmetric attacks are denoted by $\bar \Gm$ and $\bar \Gm'$.
We also define the set of twirled states as follows
\begin{definition}
The sets of all twirled bipartite states  $\bar \rho_{AB}$ with respect to the group $G$ is $$\mathbf T_G= \{ \mathcal T^{G}[\rho_{AB}] | \rho_{AB} \in \mathcal H_A \otimes \mathcal H_B \}. $$\end{definition}
The following theorem states the criteria under which two protocols have the same optimal attack. 

\begin{theorem} \label{theorem_sameattack}
If the following three conditions are satisfied, the protocols $P$ and $P'$ have the same optimal attack 

\begin{enumerate}[(I)]

\item \label{I}
$\mathbf T_{G} = \mathbf T_{G'}$.

\item \label{II}
The average measurement quantities $Q$ and $Q'$ provide the same constraints on all $\bar \rho_{AB}$ in $\mathbf T_G$ and $\mathbf T_{G'}$.
			
\item \label{III}
There exists a third group $H$ with a representation $\left \{ W_h  ; h \in H \right \}$, such that $G$ and $G'$ are subgroups of $H$, with the following properties:

\begin{enumerate}
\item \label{a} $\mathbf T_{H}=\mathbf T_{G} = \mathbf T_{G'}$, and
\item \label{b} for all POVM elements $A_{x} \in \mathbf M_A $ and $A'_{x'} \in \mathbf {M'}_{A} $ there exists a $W_{h(x,x')}$ in H such that
\begin{equation}
| \mathcal L|^2 W_{h(x,x')}^* A_{x} W_{h(x,x')}^T= | \mathcal L'|^2 A'_{x'} \label{WAx}.
\end{equation}
\end{enumerate}
Note that \ref{b} automatically implies the relation $| \mathcal L|^2W_{h(x,x')} B_{x} W^\dagger_{h(x,x')}=| \mathcal L'|^2 B'_{x'} $ for Bob's measurement operators $B_y$ defined in equation (\ref{Byproj}). 
\end{enumerate}
\end{theorem}

\startproof
We show that the optimization of the key rate in equation (\ref{rsym}) leads to the same optimal $ \rho_{AB}^{\mathrm{opt}}$ for both protocols. There are two parts to the proof. 
First, from (\ref{I}) and (\ref{II}) it follows that $\bar \Gm = \bar \Gm'$ by definition. Therefore, the set over which the key rate is optimized is identical for the two protocols. 
Second, we show that the mutual information $\bar I(\mathcal E(\bar \rho_{AB}))$ and the Holevo quantity $\bar \chi(\mathcal E(\bar \rho_{AB}))$ in equations (\ref{IONB}, \ref{chiONB}) are identical for both protocols for all states in the set $\bar \Gm$. We show this for the example of the mutual information, but the same arguments apply to the Holevo quantity as well. 

As a starting point, we switch to the language where the unitaries act on the sets $\mathbf M_A^u$ and $\mathbf M_B^u$ instead of the individual elements $A_x$ and $B_y$. The action of each unitary $U$ on the $A_x$ and $B_y$ defines uniquely the action of the same unitary $U$ on $\mathbf M_A^u$ and $\mathbf M_B^u$ by the relation in (\ref{UMU} \ref{UMUB}). This correspondence allows us to uniquely relabel $W_{h(x,x')}$ by $W_{h(u,u')}$, where $u$ and $u'$ are the basis announcements found in $P$ and $P'$, respectively. We can now restate (\ref{a}) and say that there exists unitaries $W_{h(u,u')}$ for all pairs $(u,u')$ such that 
\begin{align}
&W_{h(u,u')}^* \mathbf M_A^u W_{h(u,u')}^T = \mathbf {M'}_A^{u'}\\
&W_{h(u,u')} \mathbf M_B^u W_{h(u,u')}^\dagger = \mathbf {M'}_B^{u'}.
\end{align}

Furthermore, for a fixed $u_0$ in $P$, there exist unitaries $W_{h(u_0, u)} = W_{h(u,u')} W_{h(u',u_0)}$, connecting all $u$ in $P$ to the fixed $u_0$. Similarly, there exist unitaries $W_{h(u'_0, u')}$ connecting a fixed $u'_0$ to all $u'$ in $P'$. Using the invariance $\bar \rho_{AB} = W_h^* \otimes W_h \bar \rho_{AB} (W_h^* \otimes W_h)^\dagger$ for all $W_h \in H$ and lemma \ref{th_trafo}, the total mutual information of $P$ and $P'$ satisfies the following
\begin{align}
&\bar I(\mathcal E(\bar \rho_{AB})) =  \frac{1}{|\mathcal L|} \sum_{u=1}^{|\mathcal L|} I(\bar \rho_{AB}, \mathbf M_{AB}^u) =  I(\bar \rho_{AB}, \mathbf M_{AB}^{u_0})\nonumber, \\
&\bar I(\mathcal E'(\bar \rho_{AB})) =  \frac{1}{|\mathcal L'|} \sum_{u'=1}^{|\mathcal L'|} I(\bar \rho_{AB}, \mathbf {M'}_{AB}^{u'}) =  I(\bar \rho_{AB}, \mathbf {M'}_{AB}^{u'_0}) \nonumber.
\end{align}
A similar statement can be made for the Holevo quantity as well. 

Since there exists also a unitary $W_{h(u_0, u'_0)}$, we can conclude that
\begin{align}
I(\bar \rho_{AB}, \mathbf M_{AB}^{u_0}) =  I(\bar \rho_{AB}, \mathbf {M'}_{AB}^{u'_0})
\end{align}
from which we conclude that $\bar I(\mathcal E(\bar \rho_{AB})) = \bar I(\mathcal E'(\bar \rho_{AB}))$
Again, a similar statement holds for the Holevo quantity. 

It follows now that the same function $\bar r(\mathcal E (\rho_{AB})) = \bar r(\mathcal E' (\rho_{AB}))$ appears in the optimization of protocols $P$ and $P'$. Since these are now identical optimization problems, they must have the same solution, and therefore the same optimal attack.
\qedendproof

We will give some examples of protocols with the same optimal attack in the next section. We analyze qubit protocols, where we can make use of the point group symmetries, which are commonly used and well studied in the field of crystallography. 

\subsection{Protocols with the same optimal attack as the 6-state protocol}  \label{6-state_equiv}

The signal states of the 6-state protocol form a regular octahedron in the Bloch sphere representation. 
The symmetry group $G$ that maps an octahedron onto itself is the discrete group $O$ with a unitary irreducible representation on the qubit space. 
A symmetry group $G'$ that satisfy the criterium (\ref{I}) in theorem \ref{theorem_sameattack}
is for example the icosahedron group $I$ \cite{koster63a, butler81a}. 

\begin{figure}
\includegraphics[width=0.5\textwidth]{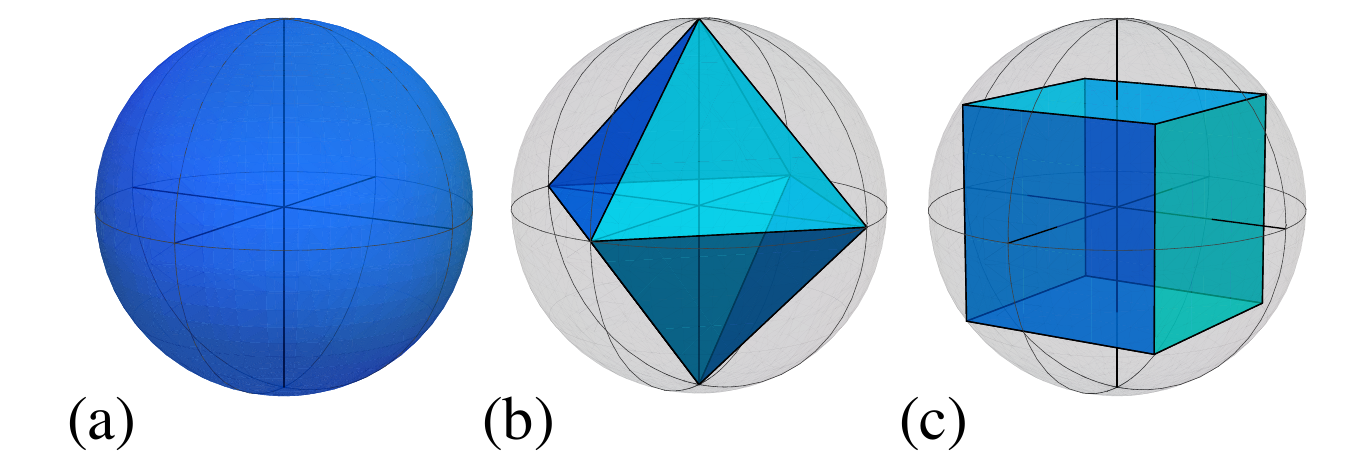}
\caption{\label{fig:SU2} Representation of sets of signal states on the Bloch sphere related to the 6-state protocol. (a) States that are invariant under $SU(2)$ (b) States of the 6-state protocol (Octahedron), invariant under O-symmetry, (c) States of the cube protocol, invariant under O-symmetry.} 
\end{figure}

We can now construct protocols with signal states that are invariant under $O$- or $I$-symmetry. 
For example, let us examine protocols, where the signal states form a cube ($O$-symmetry), a dodecahedron ($I$-symmetry) or an icosahedron (again $I$-symmetry) on the Bloch sphere,
consisting of 8, 20 and 12 states, respectively. For each state there exists an orthogonal state on the opposite side of the Bloch sphere. See Fig. \ref{fig:SU2} for a representation of the signal states of the 6-state and the cube protocol. 

The average error rate adopts the same form for all these protocols, namely, $Q= 2b$, where $b$ is defined in equation (\ref{eq_d1MUBs}). Therefore, condition (\ref{II}) in theorem \ref{theorem_sameattack} is also satisfied, and we conclude that the sets $\bar \Gm$ for these protocols are identical to the set given in the case of the 6-state protocol. 
Since there is only one state in $\bar \Gm$ for the 6-state protocol, we can already conclude that the optimal attack of the 6-state, the cube, the icosahedron and the dodecahedron protocols are the same, which was already established to be the optimal universal cloner. 

\subsection{Protocols with the same optimal attack as the BB84 protocol} \label{2n_subsubs}

\begin{figure}
\includegraphics[width=0.5\textwidth]{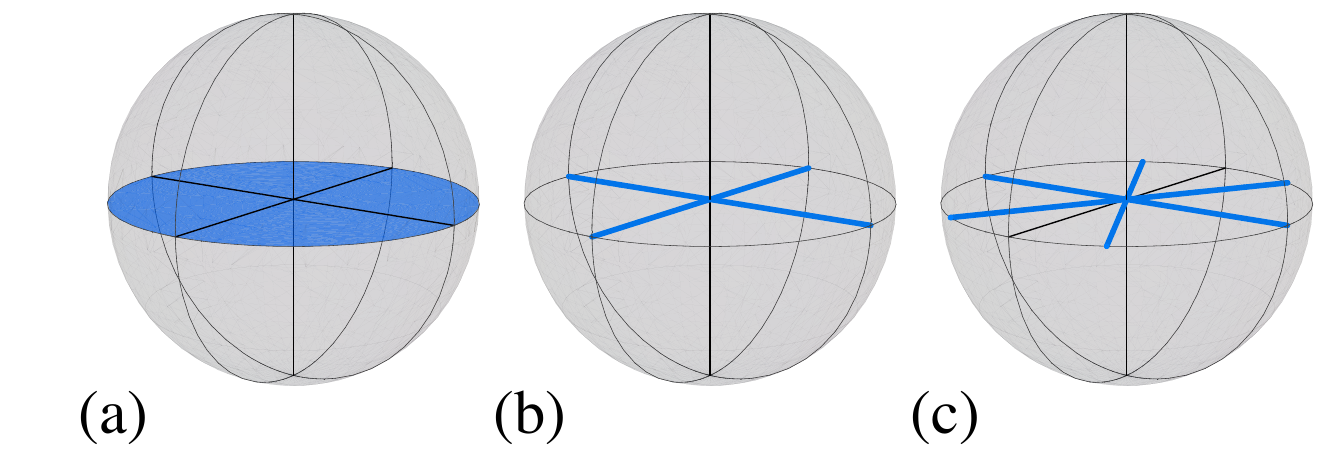}
\caption{\label{fig:cov} Representation of signal states on the Bloch sphere for protocols related to the BB84 protocol. (a) States that are invariant under $D_{\infty}$ (b) States of the BB84 protocol, $D_4$-symmetry (c) States of the $2n$ protocol for $n=3$, $D_6$-symmetry }
\end{figure}

We analyze
protocols with $2n$ $(n \geq 2)$ signal states $\ket{\varphi_x}$, that are distributed equally in the equatorial plane of the Bloch sphere, represented by the Bloch vectors
\begin{equation}
s_x^{(2n)}= ( \sin (\pi x / n)  , 0 , \cos( \pi x / n ) ) \quad x=0,...,2n-1.
\end{equation}
For each state $\ket{\varphi_x}$ there exists an orthogonal state $\ket{\bar \varphi_x}$ on the opposite side of the Bloch sphere, which together form a basis. 
For $n=2$, we recover the signal states of the BB84 protocol. In Fig. \ref{fig:cov} the signal states of the $2n$ protocols for $n=2, 3$ are represented on the Bloch sphere. 

The symmetry group of the signal states of the $2n$-protocol is called the dihedral group denoted by $D_{2n}$. 
In the multiplication tables of Refs. \cite{koster63a,butler81a}, we find the form of the set $\mathbf T_{D_{2n}}$ for $n=2,3$.
It turns out that $\mathbf T_{D_{6}} = \mathbf T_{D_{4}}$, where $\mathbf T_{D_{4}}$ contains the symmetrized states of the BB84 protocol given in equation (\ref{eq_2MUBs}) for $d=2$. 
The error rate $Q = b+c$ of the $2n$-protocol with $n=3$ is the same as for the BB84 protocol, where $b$ and $c$ are defined in equation (\ref{eq_2MUBs}). Thus we can conclude that $\bar \Gm_{6} = \bar \Gm_{4} \equiv \bar \Gm_{\mathrm{BB84}}$. 

Let us define Alice's POVM elements of the BB84 and the $2n$-protocol by $A_{x}^{(BB84)}$ for $x=1,...4$ and $A_{x'}^{(2n)}$ for $x'=1 , ... , 2n$. In both cases, the POVM elements are projectors onto the signal states.
We can identify the group $H$ in theorem \ref{theorem_sameattack} by the phase-covariant symmetry group $D_{\infty} = U(1) \times \Pi_2$, where $\Pi_2$ is the Pauli group of dimension $2$ and $U(1)$ is the unitary group.
In the tables of Ref. \cite{butler81a}, we find that the set $\mathbf T_{D_{\infty}}$ is identical to $\mathbf T_{D_4}$ and $\mathbf T_{D_{6}}$.
The phase-covariant group contains all rotations about the axis $(0, 1, 0)$ on the Bloch sphere, as well as rotations by $\pi$ about all axes lying in the $(x,z)$-plane. Thus, it also contains group elements that satisfy (\ref{b}) of theorem \ref{theorem_sameattack} for any pair $A_x^{(BB84)}$ and $A_{x'}^{(2n)}$.
Since all criteria of (\ref{III}) are satisfied by $D_{\infty}$, the optimal attack for the $2n$-protocol for $n=3$ can be identified by the optimal phase-covariant cloner, which is also the optimal attack of the BB84 protocol. 

\subsection{The cuboid protocol}  \label{cuboid_subsubs}

For some protocols with tomographically complete measurement settings the set $\bar{ \Gm}$ contains more than one state.
This has to do with loss of information during the symmetrization process. Recall that Alice and Bob only keep the averaged quantity $Q$, but otherwise ignore the measurement outcomes completely. This means that introducing symmetries to a problem can come at the expense of increasing the number of states in $\bar \Gm$. 

As an example consider a qubit protocol where the signal states lie on the corners of a rectangular cuboid. The signal states are specified by the Bloch vectors
\begin{eqnarray}
s_x=( \pm \sin \theta,  \pm \cos \theta,  0 ) \\
s_z=( 0, \pm \cos \theta , \pm \sin \theta   ),
\end{eqnarray}
where $\theta$ describes the angle between the $(0, y, 0)$-axis of the Bloch sphere and the corners of the cuboid. 
This protocol is composed of 4 bases. For each state $\ket{\varphi_x}$ there exists an orthogonal state $\ket{\bar \varphi_x}$ on the opposite side of the Bloch sphere. 

The symmetry group of this protocol is the same as of the BB84 protocol ($D_4$). Although (\ref{I}) in theorem \ref{theorem_sameattack} is satisfied, 
the error rate of the cuboid protocol is given by $Q=\frac{1}{2}(3 b + c +(b-c) \cos(2 \theta))$, which is different from the BB84 error definition. Moreover, we could not find a group $H$ to satisfy the condition (\ref{III}).
We performed numerical optimizations and found that the optimal attack of the cuboid protocol is not equal to the optimal attack on the BB84 protocol. It is also not the optimal cloner. 

Note that for $\theta =\frac{\pi}{2}$, we recover the BB84 protocol. For $\theta=\frac{\pi}{4}$ the signal states span a cube, which we already discussed in section \ref{6-state_equiv}.


\section{Conclusion} \label{conclusion_sec}

In this paper we analyze the connection between the optimal attack on a QKD protocol and the optimal cloning attack, in which the eavesdropper uses an optimal cloner to attack the protocol. 
We analyze protocols that have sufficient symmetries in the signal states and the post processing of the classical data, so that the optimal attack is, without loss of generality, a symmetric attack in the framework of the security proof of Refs. \cite{devetak05a, kraus05a}. 

We compare the optimal symmetric attack to optimal covariant cloners. 
It turns out that a necessary condition for the cloning transformation to be an optimal attack is the strong covariance condition, which guarantees that the optimal attack and the optimal cloner are chosen from the same set. 
However, this condition is not sufficient to uniquely identify the optimal attack with the optimal cloner,   
except in the case where only one state is found in the set from which the optimal attack and the optimal cloner are chosen. 
Protocols which use $d+1$ MUBs fall into this category. 

We analyze the optimal attack of protocols using $2$, $d$ and $d+1$ MUBs in $d$-dimensional Hilbert spaces. Intuitively, one expects that the optimal attack can always be identified with an optimal cloner. We prove that this intuition is correct in the case of $2$ and $d+1$ MUBs, but for protocols using $d$ MUBs, the connection between optimal attack and optimal cloner fails.

We show that two protocols using different signal states can be shown to give rise to the same optimal eavesdropping attack. Whether this is the case can be investigated by simple analysis of the symmetries of the signal states, and we give examples related to the 6-state and the BB84 protocol. 

During the preparation of this manuscript a related preprint \cite{sheridan10a} has appeared, where the key rates of protocols using $2$ and $d+1$ MUBs in $d$ dimensions were calculated. In contrast to Ref. \cite{sheridan10a}, our emphasis lies on establishing the general connection between optimal cloning and optimal eavesdropping.


\section{ Acknowledgments }

The authors would like to thank Xiongfeng Ma, Geir Ove Myhr, Tobias Moroder, Marco Piani and Volkher Scholz for useful discussions, and especially Tobias Moroder who has  provided the proof of theorem \ref{th_concavity}.  This work was supported by the European Projects SECOQC and QAP, by an NSERC Discovery Grant, OCE and Quantum Works.


\appendix


\section{Post selection} \label{post_selection_app}

In many cases Alice and Bob share data at the end of the quantum phase of their protocol, that contains unusable parts for the key generation. Typically they postselect on a set of useful data. The postselection which we describe here applies for example to basis sifting, or to the case where Alice and Bob discard data points because Bob did not receive a signal. 

Let us first examine the classical version of the postselection protocol, which starts with the ccq state $\rho_{XYE}$. In the postselection described here, an announcement is made for each individual signal.
After the quantum phase, Alice and Bob identify the weakly correlated data that they want to filter out by calculating to each measurement outcome $x$ and $y$ some values $f(x) = v$ and $f(y) = w$, and announcing $v$ and $w$ publicly. Typically, the announcements $v$ and $w$ do not reveal any information about the key. 
Based on the announcements, Alice and Bob decide of they want to keep the data or discard it. For example, they can keep only those events where $v=w \equiv u$. 
In particular, $v$ and $w$ often plays the role of a basis announcement where Alice and Bob only keep those events which they measured in the same basis, namely where $v=w \equiv u$, and discard the rest of the signals, where $v \neq w$.
By identifying the values $v$ and $w$, Alice and Bob effectively partition their original POVMs $\mathbf M_A$ and $\mathbf M_B$ into subsets $\mathbf m_A^v = \{A_x : f(x)  = v \}$ and $\mathbf m_B^w = \{B_y : f(y)  = w \}$, each containing the POVM elements labeled by the value $v$ or $w$ of the announcement. 

The quantum version of the postselection procedure is described by the map $\mathcal E$, which is composed of two consecutive steps: First, the announcement and filtering, which is jointly described by a quantum operation, and second, the measurement of the remaining states. 

The announcement and filtering is described by a quantum operation with Kraus operators $K_u \otimes L_u$ on $\mathcal H_A \otimes \mathcal H_B$. Since only events with $w = v \equiv u$ are kept, the Kraus operators come in paris $K_u \otimes L_u$ with the same index $u$.
The Kraus operators satisfy $\sum_u K_u^\dagger K_u \otimes L_u^\dagger L_u  \leq \mathbbm  1$, and they are related to the POVM elements in $\mathbf m_A^u$ and $\mathbf m_B^u$ by the rule $K_u = \sqrt{\sum_{ \mathbf m_A^u} A_{x}}$ and $L_u = \sqrt{\sum_{\mathbf m_B^u} B_{y}}$. 
The probability that the state $\rho_{AB}$ is kept during the postselection is $p_{\mathrm{kept}}$. There is also a Kraus operator corresponding to the discarded events, which happens with probability $1- p_{\mathrm{kept}}$.  
For each Kraus operator, the information $u$ is announced to all parties and stored in three classical registers $\bar A$, $\bar B$ and $\bar E$ held by Alice, Bob and Eve, respectively. 
The state after the quantum operation held by Alice, Bob and Eve, conditioned on kept events is given by 
\begin{eqnarray}
\psi=  \sum_u p(u) \ketbra{\Psi_u}{\Psi_u} \otimes \ketbra{u}{u}_{\bar A \bar B \bar E}, \label{announcement}
\end{eqnarray}
where $\ket{\Psi_u} =  K_u \otimes L_u \otimes \mathbbm 1_E \ket{\Psi}/ \sqrt{\tilde p(u)}$ is the pure state conditioned on the announcement $u$ with normalization $\tilde p(u)$, and $p(u)= \tilde p(u)/p_{\mathrm{kept}}$ is the normalized probability distribution of the announcement $u$ conditioned on events that were kept. 

The announcement and filtering step is followed by a measurement to extract the remaining data.
Each $\ket{\Psi_u}$ is measured with respect to new (normalized) POVMs $\mathbf M_A^u$ and $\mathbf M_B^u$ with elements conditioned on the announcement $u$:
\begin{align}
&\mathbf M_A^u = \{ A_{x}^u \} = \{  K_u^{-1}   A_{x}   K_u^{-1 \dagger} : A_x \in \mathbf m_A^u\}\\
& \mathbf M_B^u = \{ B_{y}^u \} = \{  L_u^{-1}  B_{y} L_u^{-1^\dagger} : B_y \in \mathbf m_B^u \}
\end{align}
The inverses $K_u^{-1}$ and $L_u^{-1}$ are defined on the non-zero subspace of $K_u$ and $L_u$ only. Again, the new POVMs only come in paris with the same index $u$. 

The measurement of $\ket{\Psi_u}$ with respect to the new POVM $\mathbf M^u = \mathbf M_A^u \otimes \mathbf M_B^u$ is equivalent to the measurement of $\ket \Psi$ with respect to the original POVM, namely $\tr_{AB} \{ A_{x}^u \otimes B_{y}^u \otimes \mathbbm 1_E \ketbra{\Psi_u}{\Psi_u} \}  = \frac{1}{\tilde p(u) } \tr_{AB} \{A_{x} \otimes B_{y} \otimes \mathbbm 1_E \ketbra{\Psi}{\Psi} \}$.
The measurement transforms $\ket{\Psi_u}$ into a ccq state 
\begin{align}
\ketbra{\Psi_u}{\Psi_u} \to \rho_{XYE}^u = \sum_{\mathbf M^u} p_u(x,y) \ketbra{x, y}{x, y}  \otimes  \rho_{E}^{x y}  \label{rhoMeasu}
\end{align}
with $p_u(x,y) = \tr \{ A_{x}^u \otimes B_{y}^u \otimes \mathbbm 1_E \ketbra{\Psi_u}{\Psi_u} \} =p(x,y) / \tilde p(u)$ and the conditional states $\rho_{E}^{xy} = \tr_{AB} \{ A_{x} \otimes B_{y} \otimes \mathbbm 1 \ketbra{\Psi}{\Psi}\} / p(x,y)$. 

We choose to calculate the key rate from each ccq state $\rho_{XYE}^u$ independently, which leads to the effective key rate
\begin{equation}
\bar r(\mathcal E(\rho_{AB})) = \sum_u p(u) r(\rho_{XYE}^u) \label{PSrtilde}.
\end{equation}


\section{Proof of the weak convexity of the classical mutual information} \label{proof_weakconvexity_app}

In this appendix we prove theorem \ref{th_weakconvexity} in section \ref{theorems_sec}. 

\startproof
The mutual information $ I(\rho_{AB},\mathbf M_{AB})$ depends only on the probability distribution $p(x,y)$. For $\bar \rho_{AB}$ the mutual information is explicitly given by $ I(\bar \rho_{AB},\mathbf M_{AB}) = H(\bar p(x)) - H(X|Y)_{\bar p}$, 
where $H(p(x)) = -\sum_x p(x) \log p(x)$ is the Shannon entropy, and $H(X|Y)_{\bar p} = \sum_{x,y} \bar p(x,y)  \log \left ( \frac{ \bar p(x,y)}{\bar  p(y)} \right )$ is the conditional entropy.
The first term satisfies 
\begin{equation}
H(\bar p(x)) = H(p(x)) = H(q(x))\label{weakshannon}, 
\end{equation}
because $p(x) = q(x) = \bar p(x)$. The second term, the conditional entropy, is concave, namely 
\begin{equation}
H(X|Y)_{\bar p} \leq    \lambda H(X|Y)_{p} + (1- \lambda) H(X|Y)_{q} \label{condentropy_concavity},
\end{equation}
where $H(X|Y)_{p}  =  - \sum_{x,y}  p(x,y)  \log \left ( \frac{ p(x,y)}{  p(y)} \right )$ and similarly for $H(X|Y)_q$. The concavity of the conditional entropy is shown by applying the log sum inequality \cite{cover06a},
\begin{align}
\left ( \sum_i a_i \right ) \log \left ( \frac{\sum_j a_j}{\sum_k b_k} \right ) \leq  \sum_i a_i \log \left (\frac{a_i}{b_i} \right ) ,
\end{align}
to $ H(X|Y)_{\bar p}$. 
Equations (\ref{weakshannon}) and (\ref{condentropy_concavity}) together imply the weak convexity of the classical mutual information. 

\qedendproof


\section{Proof of the concavity of the Holevo quantity} \label{proof_concavity_app}

In this appendix we prove theorem \ref{th_concavity} in section \ref{theorems_sec}.
We will use the traditional notation $$\chi(X:E)_{\rho_{XE}} := \sum_x p(x) S(\rho_E^x)$$ to denote the Holevo quantity of the cq state $\rho_{XE} = \sum_x p(x) \ketbra{x}{x} \otimes \rho_E^x$.

\startproof
 Given the states $\rho_{AB}$ and $\sigma_{AB}$ with purifications $\ket{\Psi}_{ABE'}$ and $\ket{\Sigma}_{ABE'}$ on the system $E= E'$. Alice measures the states with respect to the POVM elements $A_x$ and stores the result in the system $X$. 
The cq states describing the situation for Alice and Eve after the measurement are
\begin{eqnarray}
&\ket \Psi& \to  \rho_{XE'} = \sum_x p(x) \ketbra{x}{x} \otimes \rho_{E'}^{x}, \\
&\ket \Sigma  & \to  \sigma_{XE'} = \sum_x q(x) \ketbra{x}{x} \otimes \sigma_{E'}^{x} ,
\end{eqnarray}
with Eve's conditional states 
$\rho_{E'}^{x} = \tr_{AB}\{A_x \otimes \mathbbm 1 \ketbra{\Psi}{\Psi}\} / p(x)$ and $\sigma_{E'}^{x} = \tr_{AB}\{A_x \otimes \mathbbm 1 \ketbra{\Sigma}{\Sigma}\} / q(x)$. 
The Holevo quantity of $\rho_{XE'}$ and $\sigma_{XE'}$ is given by
\begin{align}
&\chi(\rho_{AB}, \mathbf M_A))= \chi(X:E')_{\rho_{XE'}} \nonumber,\\ 
&\chi(\sigma_{AB}, \mathbf M_A))= \chi(X:E')_{\sigma_{XE'}} \label{chi_rho_sigma}. 
\end{align}

We construct a particular purification on the joint system $E=E'F$ for the convex sum $\bar \rho_{AB} = \lambda \rho_{AB} + (1-\lambda) \sigma_{AB}$:
\begin{equation}
\ket{\bar \Psi}_{ABE'F}=\sqrt{\lambda} \ket{\Psi}\ket{0}_{F}+ \sqrt{1-\lambda} \ket{\Sigma}\ket{1}_{F}.
\end{equation}
After measuring $\ket{\bar \Psi}_{ABE'F}$ with respect to $A_x$, the state shared by Alice and Eve is
\begin{align}
\bar \rho_{XE'F} = \sum_x \bar p(x) \ketbra{x}{x} \otimes \bar \rho_{E'F}^{x} 
\end{align}
with Eve's conditional states $\bar \rho_{E'F}^{x} = \tr_{AB}\{A_x \otimes \mathbbm 1_{BE'F} \ketbra{\bar \Psi}{\bar \Psi}\} / \bar p(x) $ and $\bar p(x) = \lambda p(x) +(1-\lambda) q(x)$. The Holevo quantity of $\bar \rho_{XE'F}$ is
\begin{align}
\chi(\bar \rho_{AB}, \mathbf M_A)) =  \chi(X:E'F)_{\rho_{XE'F}}\nonumber. 
\end{align}

Let $\mathcal M: F \to F'$ be a trace-preserving quantum operation. 
For our purposes, we identify $\mathcal M$ with a measurement on $F$ in the standard basis $\{\ket 0 , \ket 1\}$ and write the outcome in a new register $F'$. By defining $\lambda_x = \lambda \frac{p(x)}{\bar p(x)}$, the state after the measurement is given by
\begin{eqnarray}
&\rho_{XE'F'}  = & \sum_x \bar p(x) \ketbra{x}{x} \otimes [\lambda_x  \rho_{E'}^{x} \otimes \ketbra{0}{0}_{F'}  \\
&&+ (1-\lambda_x) \sigma_{E'}^{x} \otimes \ketbra{1}{1}_{F'}].
\end{eqnarray}

Let us state a lemma about the Holevo quantity extracted from a state of the form $\rho_{XE'F'}$:
\begin{lemma} \label{lemma_ent_conc}
The Holevo quantity extracted from $\rho_{XE'F'}$ satisfies
\begin{eqnarray}
&\chi(X:E'F')_{\rho_{XE'F'} } \geq& \lambda \chi(X:E')_{\rho_{XE'}} \label{chicon} + \\
&&(1-\lambda) \chi(X:E')_{\sigma_{XE'}}\nonumber,
\end{eqnarray}
with $\chi(X:E')_{\rho_{XE'}} $ and $\chi(X:E')_{\sigma_{XE'}}$ given in equation (\ref{chi_rho_sigma}).
Equality holds, if the probabilities $\lambda_x$ are independent of $x$, $\lambda_x = \lambda $. 
\end{lemma}
\startproof
From $\rho_{XE'F'}$ we calculate the Holevo quantity using the joint entropy theorem of the von Neuman entropy \cite{nielsen00a} 
\begin{eqnarray}
&\chi(X:EF') & =   \lambda \chi(X:E')_{\rho_{XE'}} + (1-\lambda) \chi(X:E')_{\sigma_{XE'}}\nonumber \\
&&+ h(\lambda) - \sum_x \bar p(x) h(\lambda_x)\label{chiXEF'},
\end{eqnarray}
with the binary entropy function $h(x) = -x\log(x) - (1-x)\log(1-x)$, and where $\chi(X:E')_{\rho_{XE'}} $ and $\chi(X:E')_{\sigma_{XE'}}$ are given in equation (\ref{chi_rho_sigma}).
From the concavity of the Shannon entropy it follows that $h(\lambda) - \sum_x \bar p(x) h(\lambda_x) \geq 0$, and in particular, if $\lambda = \lambda_x$ for all $x$, the equality holds. 
\qedendproof

According to \cite{nielsen00a}, a map of the form $\mathcal M: F \to F'$ can only decrease the Holevo quantity:
\begin{equation}
\chi(X:E'F)_{\rho_{XE'F}} \geq \chi(X:E'F')_{\rho_{XE'F'}} \label{decreaseinfo}.
\end{equation}
Equation (\ref{chicon}) together with equation (\ref{decreaseinfo}), show that the desired result. 
\qedendproof


\section{Proof of lemma \ref{lemma_Gstarinv}} \label{Ax_app_sec}

In this appendix we prove the $G^*$-invariance of $\rho_A$ and $A_x$ for the source-replacement scheme in section \ref{source_sec}, if $p(x)$ is uniformly distributed and if the signal states are $G$-invariant. 

\startproof
We trace out system $S$ from the source state $\ket \Phi_{AS}$ in equation (\ref{compressed_source}) and identify Alice's reduced state $\rho_A$ by
\begin{equation}
\rho_A = \sum_i \kappa_i \ketbra{i}{i} = \sum_x p(x) \ketbra{\varphi_x}{\varphi_x} \label{phi_signals}.
\end{equation}
From the $G$-invariance of the signal states and the uniform distribution of $p(x)$, it follows that $\rho_A$ is also $G$-invariant. 
Since $\rho_A$ is diagonal in the basis $\mathcal B$ with real eigenvalues $\kappa_i$, it holds that $$\rho_A^*= \rho_A.$$ Therefore, $\rho_A$ is also $G^*$-invariance (where the complex conjugate is taken with respect to the Schmidt basis) as can be easily seen from $U_g^* \rho_A U_g^T = (U_g \rho_A U_g^\dagger )^*= \rho_A$.

Due to the positivity of the coefficients $\kappa_i$ and the full rank of $\rho_A$, the square root $\sqrt{ \rho_A}$ and the inverse $\rho_A^{-1}$ are well-defined. The $G$- and $G^*$-invariance of $\rho_A^{-1}$ can be straightforwardly verified: $U_g \rho_A^{-1} U_g^\dagger = (U_g \rho_A U_g^\dagger )^{-1} = \rho_A^{-1} $, and similarly for the $G^*$-invariance. 

In \cite{christandl07a}, it is shown for permutation groups, that for every positive $G$-invariant operator $\rho_A$, $\sqrt{\rho_A}$ is also $G$-invariant. The same proof applies also here, and thus, the $G$- and $G^*$-invariance of $\rho_A$ also implies the $G$- and $G^*$-invariance of $\sqrt{\rho_A}$. 

By using the $G$-invariance of the signals states, and the $G^*$-invariance of $\sqrt{\rho_A}$ and $\rho_A^{-1}$, the $G^*$-invariance of $A_x$ follows directly from the definition in equation (\ref{Ax}).


\section{Properties of symmetrized states}\label{Schurs_lemma_app}

The symmetrized states with the commutation property (\ref{commutator}) can be characterized using Schur's lemma.
Let the unitaries $U_g^{(\mu)}$ denote the irreducible representation $\mu$ of a group $G$ on the Hilbert spaces $\mathcal H_\mu$.
Every reducible representation of $G$ can be decomposed into a direct sum of irreducible representations.
For example, the tensor product $U_g^{(\nu)*} \otimes U_g^{(\nu)} $ is a reducible representation of $G$ and can be decomposed into a (block diagonal) direct sum of irreducible representations $U_{g,i}^{(\mu)}$ defined on the spaces $\mathcal H_{ i}^{(\mu)}$ carrying the irreducible representations $\mu$
\begin{equation}
U_g^{(\nu)*} \otimes U_g^{(\nu)} =  \bigoplus_\mu \bigoplus_{i=1}^{m_\mu}  U_{g,i}^{(\mu)}  .
\end{equation}
Each irreducible representation $\mu$ occurs with integer multiplicity $m_\mu$ indicated by the index i. 

According to Schur's lemma, an operator $\tau_{i,j}^{(\mu, \nu)}: \mathcal H_{ i}^{(\mu)} \to \mathcal H_{ j}^{(\nu)}$ that satisfies $\tau_{i,j}^{(\mu, \nu)} U_{g,i}^{(\mu)} =U_{g,j}^{(\nu)}  \tau_{i,j}^{(\mu,\nu)}$ for all $g \in G$ is either (i) the identity map from $\mathcal H_{i}^{(\mu)} \to \mathcal H_{j}^{(\mu)}$ if $\mu = \nu$, or (ii) equal to zero. 
Using Schur's lemma, every positive operator $\bar \rho_{AB}$ that commutes with all the group elements of $U_g^{(\nu)*} \otimes U_g^{(\nu)}$ is characterized by
\begin{equation}
\bar \rho_{AB} = \bigoplus_\mu \bigoplus_{i,j=1}^{m_\mu} c_{ij}^{(\mu)} \tau_{ij}^{(\mu)} 
\end{equation}
where the $c^{(\mu)}$ are $m_\mu \times m_\mu$ matrices with positive eigenvalues $\lambda_i^{(\mu)}$ and $\tau_{ij}^{(\mu)}$ is the identity map between the subspaces with equal representation. 
After diagonalizing $c^{(\mu)}$, we can rewrite $\bar \rho_{AB}$ in a block-diagonal form on a new decomposition $\mathcal K_i^{(\mu)}$ of the Hilbert space by
\begin{equation}
\bar \rho_{AB}=\bigoplus_\mu \bigoplus_{i=1}^{m_\mu} \lambda_i^{(\mu)} \mathbbm 1_i^{(\mu)} \label{decomp}
\end{equation}
where $\mathbbm 1_i^{(\mu)}$ is the identity on the subspace $\mathcal K_i^{(\mu)}$.
The block-diagonal form of $U_g^{(\nu)*} \otimes  U_g^{(\nu)}$ is quasi ``inherited" by $\bar \rho_{AB}$.


\section{Post selection on orthonormal bases}\label{ONB_PS_app}

In this appendix we calculate the filters $K_u$ and $L_u$ for the class of protocols described in section \ref{ONB_sec}.

Given a protocol where the set of signal states $\mathbf S_{\mathcal L}$ contains $|\mathcal L|$ complete bases. Alice and Bob announce the basis of each signal, which partitions the POVMs $\mathbf M_A$ and $\mathbf M_B$ into $|\mathcal L|$ disjoint sets $\mathbf m_A^v = \{A_{(\beta,k)} : \beta = v \}$ and $\mathbf m_B^w = \{B_{(\beta,k)} : \beta = w \}$. They decide to keep only those events which were measured in the same basis, e.g.,those where they made the same announcement $f(x) = f(y) = u$. Each set is associate with the filters
\begin{align}
K_u = L_u = \frac{\mathbbm 1}{\sqrt{|\mathcal L|}},
\end{align}
which are proportional to the identity for all $u$.

Let Alice and Bob share a state $\rho_{AB}$ with a purification $\ket \Psi$. Due to the particular form of the filters, it follows that $\ket{\Psi_u} = \ket \Psi$, and that the new POVMs $\mathbf M_A^u$ and $\mathbf M_B^u$ are only a rescaled version of the old ones 
\begin{align}
\mathbf M_A^u = \{ |\mathcal L| \; A_{(\beta, k)} : \beta = u \} \\
\mathbf M_B^u = \{ |\mathcal L| \; B_{(\beta, k)} : \beta = u \} 
\end{align}

\bibliographystyle{unsrt}
\bibliography{qit_20090708,myown}

\end{document}